\documentclass[aps,physrev,amsmath,amssymb,amsfonts,lengthcheck,longbibliography,floatfix,superscriptaddress]{revtex4-2}

\usepackage[bbgreekl]{mathbbol}
\usepackage{color, graphicx}\graphicspath{ {figures/} }
\definecolor{linkcolor}{rgb}{0,0,0.6}
\usepackage[pdftex, colorlinks=true,
	pdfstartview = FitV,
	linkcolor    = linkcolor,
	citecolor    = linkcolor,
	urlcolor     = linkcolor,	
	hyperindex   = true,
	hyperfigures = false]{hyperref}

\newcommand{\pd}{\partial}
\newcommand{\avg}[1]{\langle  #1 \rangle}
\newcommand{\Avg}[1]{\left\langle  #1 \right\rangle}

\newcommand{\bs}{\boldsymbol}
\newcommand{\bb}{\mathbb}

\newcommand{\dfn}{=}
\newcommand{\trp}{\intercal}

\newcommand{\dif}{\nabla}
\newcommand{\pos}{{\bf r}}

\newcommand{\crt}{J}
\newcommand{\mob}{\lambda}
\newcommand{\frc}{F}
\newcommand{\cpl}{C}
\newcommand{\spu}{\nu}
\newcommand{\noi}{\Lambda}

\newcommand{\fld}{\phi}
\newcommand{\fue}{n}
\newcommand{\act}{\Delta\mu}
\newcommand{\D}{\mathrm d}

\begin{document}

\title{Control of active field theories at minimal dissipation}

\author{Artur Soriani}
\affiliation{School of Mechanical Engineering, Tel Aviv University, Tel Aviv 69978, Israel}
\affiliation{Center for Physics and Chemistry of Living Systems, Tel Aviv University, Tel Aviv 69978, Israel}

\author{Elsen Tjhung}
\affiliation{School of Mathematics and Statistics, The Open University, Walton Hall, Milton Keynes, MK7 6AA, United Kingdom}

\author{Étienne Fodor}
\affiliation{Department of Physics and Materials Science, University of Luxembourg, L-1511 Luxembourg City, Luxembourg}

\author{Tomer Markovich}
\affiliation{School of Mechanical Engineering, Tel Aviv University, Tel Aviv 69978, Israel}
\affiliation{Center for Physics and Chemistry of Living Systems, Tel Aviv University, Tel Aviv 69978, Israel}

\date{\today}

\begin{abstract}

Advances in experimental techniques enable the precise manipulation of a large variety of active systems, which constantly dissipate energy to sustain nonequilibrium phenomena without any equilibrium equivalent. To design novel materials out of active systems, an outstanding challenge is to rationalize how material properties can be optimally controlled by applying external perturbations. However, equilibrium thermodynamics is inadequate to guide the control of such nonequilibrium systems. Therefore, there is a dire need for a novel framework to provide a systematic toolbox for the thermodynamic control of active matter. Here, we build an optimization procedure for generic active field theories within a thermodynamically consistent formulation. Central to our approach is the distinction between the \textit{protocol heat}, which is dissipated only during manipulation, and the \textit{total heat}, which also accounts for the post-manipulation dissipation. We demonstrate that the latter generically features a global minimum with respect to the protocol duration. We deploy our versatile approach to an active theory of phase separation, and examine the scalings of the optimal protocol duration with respect to activity and system size. Remarkably, we reveal that the landscape of steady-state dissipation regulates the crossover between optimal control strategies for a finite duration.

\end{abstract}

\maketitle


\section*{Introduction}\label{sec:Intro}

Equilibrium thermodynamics states that manipulating passive systems at minimal dissipation can only be achieved in the quasistatic regime of infinitely slow perturbation~\cite{Jarzynski2011}. Active systems evade such a constraint due to the constant consumption of energy at the basis of their internal sustained motion and/or mechanical stress~\cite{Marchetti2013, Bechinger2016}. Then, the dissipation resulting from manipulating active systems can exhibit a minimum at a finite protocol duration~\cite{Davis2024}. This behaviour defies conventional intuition, and requires a novel optimization approach beyond that of passive systems.

Experimental techniques have demonstrated the ability to reliably manipulate active systems; for instance, using magnetic fields~\cite{Guillamat2016, Matsunaga2019}, light sources~\cite{Palacci2013, Shin2017, Maggi2018, zhang2021, nishiyama2024}, or nematic pumps~\cite{velezceron2024}. These developments call for a theoretical framework to guide experiments from empirical to optimal control. The methods of stochastic thermodynamics set the basis for building such a framework~\cite{Seifert2012}. These methods have mostly been deployed in systems with a few degrees of freedom~\cite{Schmiedl2007, Blaber2023, Loos2024}, although some studies have considered controlling equilibrium spin models~\cite{Rotskoff2015, Gingrich2016, Rotskoff2017}. To control active systems~\cite{Alvarado2025}, previous theoretical studies have offered some strategies for the navigation~\cite{Schneider2019, Liebchen2019, Piro2021, sinha2023} and confinement~\cite{Baldovin2023, Garcia2024, wang2024, Schuttler2025} of particles, some of which using machine learning~\cite{Falk2021, Casert2024}. Other studies have addressed the control of hydrodynamic theories describing active matter as continuum materials~\cite{Norton2020, Shankar2022, ghosh2024b, ghosh2024, krishnan2024}. Remarkably, most control strategies put forward so far at the hydrodynamic level neglect the role of noise, which hinders their use in active systems where fluctuations are predominant.

The collective behaviors in passive systems have been successfully studied using hydrodynamic field theories~\cite{ChaikinLubensky1995}. Similarly, nonequilibrium field theories provide a broad framework to describe the large-scale features of active systems~\cite{Marchetti2013, Marchetti2016, Markovich2021}, such as motility-induced phase separation~\cite{Cates2015}, collective directed motion~\cite{Chate2020}, and odd viscosity~\cite{MarLub2021, Fruchart2023, MarLub2024}. In particular, field theories delineate distinct classes of systems based on the symmetries of their microscopic interaction. Most of these theories are built to capture some nonequilibrium collective behaviours without any energetic consideration. Instead, recent studies have proposed thermodynamically consistent formulations which entail unambiguous quantification of dissipation~\cite{Pietzonka2018, Gaspard2019, Markovich2021, Datta2022, Chatzittofi2024, Bebon2024, Agranov2024, Sorkin2024} by describing how active systems couple with equilibrium reservoirs~\cite{Seifert2012, Fodor2022, Aslyamov2023, Falasco2024}.

\begin{figure*}
\includegraphics[width=\linewidth]{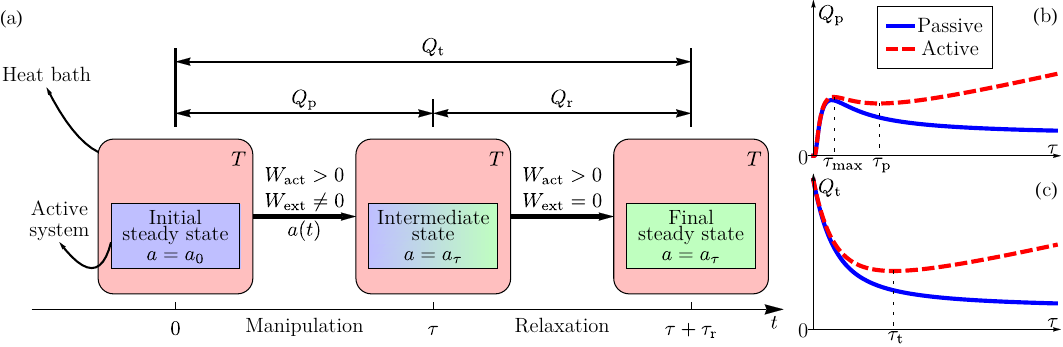}
	\caption{(a)~During manipulation ($0<t<\tau$), the system changes its state under the protocol $a(t)$: the system  accumulates external work ($W_{\rm ext}$), and dissipates protocol heat ($Q_{\rm p}$) into the thermostat. During relaxation ($\tau<t<\tau+\tau_{\rm r}$), the system changes state at constant $a(t)=a_\tau$: there is no external work, and the system dissipates relaxation heat ($Q_{\rm r}$). The active work $W_{\rm act}>0$ is produced during both manipulation and relaxation.
	(b)~The protocol heat $Q_{\rm p}$ vanishes for instantaneous manipulation ($\tau=0$), and can feature a local maximum at a finite duration ($\tau=\tau_{\rm max}$). For passive systems, $Q_{\rm p}$ saturates at large $\tau$; for active systems, $Q_{\rm p}$ diverges at large $\tau$ due to internal activity, and can feature a local minimum ($\tau=\tau_{\rm p}$).
	(c)~The total heat $Q_{\rm t} = Q_{\rm p} + Q_{\rm r}$ has a non-zero value at $\tau=0$. For passive systems, $Q_{\rm t}$ decreases towards a finite value; for active systems, $Q_{\rm t}$ diverges at large $\tau$, and {\it always} features a local minimum ($\tau=\tau_{\rm t}$).
	}
	\label{fig:ProcessHeatVsTotalHeat}
\end{figure*}

In this paper, we search for generic control strategies in field theories of active matter. We propose a versatile framework to predict the optimal protocols which change material properties by tuning some experimentally accessible parameters. Inspired by a previous optimization procedure, originally proposed for passive systems~\cite{Sivak2012} and recently adapted to active matter~\cite{Davis2024}, we focus on minimizing the heat generated by smooth and slow protocols. In contrast with~\cite{Sivak2012, Davis2024}, we argue that the \textit{protocol heat} $Q_{\rm p}$, which is dissipated only during the protocol, is less adequate for optimization than the \textit{total heat} $Q_{\rm t}$, which also accounts for the post-protocol dissipation [Fig.~\ref{fig:ProcessHeatVsTotalHeat}(a)]. For passive systems, $Q_{\rm t}$ is optimal at long protocol duration, as expected from standard thermodynamics~\cite{Jarzynski2011}, while it reaches a minimum at a finite duration for active systems [Fig.~\ref{fig:ProcessHeatVsTotalHeat}(b)]. Instead, $Q_{\rm p}$ is always zero for a vanishing protocol duration, and need not feature any other local minimum even for active systems [Fig.~\ref{fig:ProcessHeatVsTotalHeat}(c)].

We apply our framework to a thermodynamically consistent field theory, referred to as Chemical Model B (CMB), which describes nonequilibrium phase separation. Using dynamical response theory, we provide analytical predictions for protocols driving some homogeneous configurations of the system, and examine the scalings of the optimal duration with respect to system size and activity. Remarkably, we reveal a crossover between control strategies, and rationalize it based on the landscape of steady-state dissipation. Overall, our results demonstrate the potential of our novel approach to guiding control in a large class of active systems.


\section*{Results}

\subsection*{The role of relaxation: Protocol vs total heat}

We consider the external manipulation of a system at constant temperature $T$ through a finite-time protocol varying the parameter $a(t)$ from $a_0=a(0)$ to $a_\tau=a(\tau)$. The system is assumed to be initially ($t=0$) at steady state [Fig.~\ref{fig:ProcessHeatVsTotalHeat}(a)]. For any finite protocol duration $\tau$, the protocol is not quasi-static, so that the system is not in steady state at any time $t\in(0,\tau)$. After the protocol ($t>\tau$), the parameter $a(t>\tau)=a_\tau$ is constant, and the system relaxes towards the stationary configuration given by $a_\tau$. We define the post-protocol time $\tau_{\rm r}$, which is far larger than the relaxation of the slowest mode, such that the system reaches a steady state at time $\tau+\tau_{\rm r}$.

We denote by $Q_{\rm p}$ the \textit{protocol heat} dissipated by the system during the protocol. During the subsequent relaxation, an additional contribution, referred to as the \textit{relaxation heat} $Q_{\rm r}$ is dissipated, although the parameter $a(t>\tau)=a_\tau$ is constant. The \textit{total heat} $Q_{\rm t}$, which is dissipated during the protocol and the post-protocol relaxation, follows as $Q_{\rm t} = Q_{\rm p} + Q_{\rm r}$. Clearly, $(Q_{\rm p}, Q_{\rm r}, Q_{\rm t})$ all strongly depend on the shape of the protocol $a(t)$ and its duration $\tau$. Conservation of energy states that $Q_{\rm t}$ is given by the sum of the total work $W_{\rm t}$ and the change in the system energy $E$:
\begin{equation}\label{eq:TotalHeat}
	Q_{\rm t} = W_{\rm t} + E(t=0) - E(t={\tau+\tau_{\rm r}}) \, ,
\end{equation}
where $W_{\rm t}$ can be decomposed in terms of the contributions produced by the external manipulation ($W_{\rm ext}$, accumulated only during the protocol) and by any non-conservative active forces ($W_{\rm act}$, accumulated during both the protocol and the subsequent relaxation):
\begin{equation}\label{eq:work}
	W_{\rm t} = W_\text{ext} + W_\text{act} \, .
\end{equation}
In the absence of external manipulation, the quantity $W_\text{act}$ is the energetic cost needed to sustain non-equilibrium steady states \cite{Hatano2001}.
In the presence of fluctuations, Eqs.~\eqref{eq:TotalHeat} and~\eqref{eq:work} take the same form for the stochastic definitions of heat, work, and energy~\cite{Seifert2012, Fodor2022}. In what follows, we focus on average values; in particular, the averages of $E$ at times $t=0$ and $t=\tau+\tau_{\rm r}$ are evaluated in steady state, and thus are independent of the protocol.

For passive systems ($W_\text{act}=0$), Eq.~\eqref{eq:TotalHeat} reduces to the first law of thermodynamics in its standard form, so minimizing either $Q_{\rm t}$ or $W_{\rm ext}$ leads to the same optimal protocol. The second law of thermodynamics states that $W_{\rm ext}$ can only be larger than the difference between the Helmholtz free energy evaluated at $a_0$ and $a_\tau$~\cite{Jarzynski2011}. Likewise, $Q_{\rm t}$ is bounded from below by the difference of the system entropy at $a_0$ and $a_\tau$, and the bound is only saturated for quasi-static protocols. For instantaneous protocols ($\tau = 0$), $Q_{\rm p}$ vanishes since there is no sufficient time to dissipate heat during the protocol [Fig.~\ref{fig:ProcessHeatVsTotalHeat}(b)], whereas $Q_{\rm t}$ has a non-zero value as a signature of the dissipation ensued during the post-protocol relaxation [Fig.~\ref{fig:ProcessHeatVsTotalHeat}(c)].

The situation is markedly different for active systems. In steady state, $W_\text{act}$ increases extensively with time to sustain the internal processes at the basis of activity. Then, at large duration $\tau$, both $Q_{\rm t}$ and $Q_{\rm p}$ linearly increase with $\tau$, in contrast with passive systems, whereas $W_{\rm ext}$ saturates as in the passive case. It follows that $Q_{\rm t}$ {\it always} features a global minimum at finite $\tau$ [Fig.~\ref{fig:ProcessHeatVsTotalHeat}(c)], which achieves the best trade-off between the dissipation coming from internal processes and that stemming from external manipulation. Remarkably, $Q_{\rm p}$ vanishes at $\tau=0$ as in passive systems, and it may feature a local minimum $\tau_{\rm p}$ under certain conditions [Fig.~\ref{fig:ProcessHeatVsTotalHeat}(b)], in contrast with passive systems. For instance, for a scalar field theory, these conditions require a small activity (and also small system sizes, if the field is conserved), as we discuss in detail in the next sections.

In this paper, minimizing $Q_{\rm t}$ (rather than $W_\text{ext}$~\cite{Sivak2023} or $Q_{\rm p}$~\cite{Davis2024}) is the objective of our optimization. In particular, our aim is to quantify the optimal duration $\tau_{\rm t}$ at minimal $Q_{\rm t}$, which separates the two regimes of dissipation inherent to active systems [Fig.~\ref{fig:ProcessHeatVsTotalHeat}(b)]. Specifically, we deploy a systematic optimization framework~\cite{Sivak2012, Davis2024} in a thermodynamically consistent formulation of active field theories~\cite{Markovich2021}, examine the corresponding scalings of $\tau_{\rm t}$, and unveil a crossover between optimal protocols.


\subsection*{Energetics of active field theories}

We describe the energy exchange between a given active system and its surrounding reservoirs within linear irreversible thermodynamics~\cite{Mazur, Markovich2021}; the energy reservoir absorbs the heat dissipated by the system, and the internal active driving is sustained by some active driving reservoir (typically chemical fuel). In practice, we consider here the coupling of two scalar fields: the density of active particles $\fld(\pos,t)$, and the concentration of fuel molecules $\fue(\pos,t)$. The specific nature of  active driving is unimportant~\footnote{What is important is the time-reversal signature of the active driving. For instance, a flux of photons has a different time-reversal signature than chemical driving~\cite{Markovich2021}.}, and we focus on the case of chemical reactions for illustration purposes. The interactions between active particles are described by the free energy functional $F([\fld],a)$, which should not be confused with the Helmholtz free energy bounding $W_{\rm ext}$ in passive systems~\cite{Jarzynski2011}. The system is maintained away from equilibrium by setting the difference of some chemical potentials $\Delta\mu$ constant; for instance, involving reactant and product molecules whose reaction fuels the particle activity~\cite{Markovich2021}.

Given that the total number of active particles is conserved, $\phi$ obeys a conservation law in terms of the diffusive current ${\bs J}$:
\begin{equation}\label{eq:cons}
	\dot\fld + \bs\dif \cdot \bs\crt = 0 \, ,
\end{equation}
where $\dot\fld = \pd_t \fld$.
We describe the stochastic coupling of $(\phi,n)$ through a linear relation between thermodynamic fluxes $\bb\crt = (\bs\crt,\dot\fue)$ and forces $\bb\frc =(-\bs\dif (\delta F/\delta\fld), \act )$~\cite{Markovich2021}:
\begin{equation}\label{eq:MatrixDynamics}
\bb\crt = \bb L \cdot \bb\frc + T^{1/2} \bb\noi + T \bbnu \,.
\end{equation}
where the noise term $\bb\noi$ is Gaussian with zero mean and correlations given by:
\begin{equation}\label{eq:noise}
\Avg{ \bb\noi(\pos,t) \bb\noi^\trp(\pos',t') } = 2 \bb L\delta(\pos-\pos') \delta(t-t') \, .
\end{equation}
Here, $^\trp$ refers to vector transpose, and we have set the Boltzmann constant to unity. Whenever $\bb L$ depends on $\phi$ and its gradients, the noise $\bb\noi$ is multiplicative. Then, interpreting Eq.~\eqref{eq:MatrixDynamics} within the Stratonovitch convention~\cite{Gardiner, Cates2022}, the spurious drift $\bbnu$ can be explicitly calculated for any given $\bb L$ (see Ref.~\cite{SM}, section VI).

For $d$ spatial dimensions, $\bb\crt$ and $\bb\frc$ are vectors in $d+1$ dimensions. The Onsager matrix $\bb L$ is symmetric and semi-positive definite:
\begin{equation}\label{eq:OnsagerMatrix}
\bb L = \begin{pmatrix}
\mob_\fld \bs 1_d  & \bs\cpl(\fld,\bs\dif\fld) \\
\bs\cpl^\trp(\fld,\bs\dif\fld) & \mob_n
\end{pmatrix}\,, 
\end{equation}
where $(\mob_\fld,\mob_\fue)$ are mobility constants and $\bs 1_d$ is the $d\times d$ identity matrix. Physically, $\delta F/\delta\phi$ represents the free-energy cost of adding a particle, while $\Delta\mu$ corresponds to the energy gain per fuel molecule during the chemical reaction. The diagonal terms in Eq.~(\ref{eq:OnsagerMatrix}) indicate that the negative gradient of $\delta F/\delta\phi$ drives the diffusive current $\bs\crt$, while $\Delta\mu$ drives the chemical reaction rate $\dot{\fue}$. The off-diagonal terms in Eq.~(\ref{eq:OnsagerMatrix}) couple the chemical reaction to the diffusive dynamics. In general, the coupling term $\bs\cpl$ is a function of $\fld$ and its gradients. Through this coupling, $\Delta\mu$ indirectly drives $\phi$ to operate away from equilibrium.

Considering a given protocol $a(t)$ [Fig.~\ref{fig:ProcessHeatVsTotalHeat}], the expression of $Q_{\rm t}$ follows from the product between the thermodynamic fluxes and forces~\cite{Fodor2022}:
\begin{equation}\label{eq:heat}
	Q_{\rm t} = \Avg{ \int_0^{\tau+\tau_{\rm r}} \int_V \bb J \cdot \bb F \, \D{\bf r} \D t } \,,
\end{equation}
where $\langle\cdot\rangle$ denotes an average over noise realizations, and $V$ refers to the system size. The conservation of energy [Eq.~\eqref{eq:TotalHeat}] relates $Q_{\rm t}$ to the variation of the system energy $E= \Avg{F}$ and to the total work $W_{\rm t}$; see stochastic thermodynamics in Methods. The work produced by externally varying $a(t)$ during the protocol reads
\begin{equation}\label{eq:ExternalWork}
	W_\text{ext} = \int_0^{\tau} \dot a \Avg{ \pd_a F } \D t \, .
\end{equation}
The work produced by the chemical reaction over the entire process is given by
\begin{equation}\label{eq:ActiveWork_0}
	W_\text{act} = \Avg{ \int_0^{\tau+\tau_{\rm r}} \int_V \Delta\mu \dot n \, \D{\bf r} \D t } \, ,
\end{equation}
which can also be written as
\begin{equation}\label{eq:ActiveWork}
	W_\text{act} = (\tau+\tau_{\rm r}) P_0 + \int_0^{\tau+\tau_{\rm r}} \avg{P} \D t \, ,
\end{equation}
where the power $P_0 = \mob_\fue V \act^2$ is a background contribution, independent of the coupling between the diffusive and chemical sectors of the dynamics, and 
\begin{equation}\label{eq:ActivePowerDefinition}
\begin{aligned}
	P = \int_V p \, \D{\bf r} \, ,\quad
	p \dfn \frac{\act}{\mob_\fld} \bs\cpl \cdot \left( \bs\crt - \act\, \bs\cpl \right)
\end{aligned}
\end{equation}
embodies how the $\phi$-dynamics affects the dissipation rate. It is worth noting that $W_\text{act}\geq0$, as discussed in~\cite{Markovich2021}. The expressions of heat and work [Eqs.~(\ref{eq:heat}-\ref{eq:ActivePowerDefinition})] straightforwardly extend to more complex field theories~\cite{Markovich2021, Fodor2022}, beyond the simple case of a single conserved scalar field [Eq.~\eqref{eq:cons}].

The relation between the protocol heat $Q_{\rm p}$, the energy $E$, and the total work $W_{\rm t}$ is identical to that for $Q_{\rm t}$ in Eq.~\eqref{eq:TotalHeat}, with two key modifications: (i)~$E(\tau+\tau_{\rm r})$ is replaced by $E(\tau)$, where $E(\tau)$ denotes a non-stationary average of $F$ at time $\tau$; and (ii)~in the definition of $W_{\rm act}$ [Eq.~\eqref{eq:ActiveWork}], the time $\tau+\tau_{\rm r}$ is replaced by $\tau$.

Importantly, the averages in Eqs.~(\ref{eq:ExternalWork}-\ref{eq:ActivePowerDefinition}) are not stationary, and depend on the details of the protocol $a(t)$. Whenever $a$ is fixed ($\dot a=0$ and $W_{\rm ext}=0$), these averages reduce to their steady-state values, so that $W_{\rm act}$ scales like $\tau+\tau_{\rm r}$. In such a case, we have $E(0)=E(\tau)=E(\tau+\tau_{\rm r})$, and we deduce from Eq.~\eqref{eq:TotalHeat} that both $(Q_{\rm p}, Q_{\rm t})$ diverge with the protocol duration. This regime reproduces the scaling of slow protocols [Figs.~\ref{fig:ProcessHeatVsTotalHeat}(b-c)], and is a direct consequence of the permanent fuel consumption, which is present even when the parameter $a$ is kept constant.


\subsection*{Perturbative treatment for slow protocols}

We offer a systematic approach to express the heats $(Q_{\rm p},Q_{\rm t})$ as functionals of the protocol $a(t)$ for a generic active field theory [Eqs.~(\ref{eq:cons}-\ref{eq:OnsagerMatrix})]. Inspired by previous works~\cite{Sivak2012, Davis2024}, we rely on a perturbative expansion around the quasistatic limit ($\Omega\tau\gg 1$, where $\Omega$ is the slowest relaxation frequency of the system (see Ref.~\cite{SM}, section IV.B), and assume that the protocol is smooth by neglecting any discontinuity in $a(t)$; see dynamical response theory (DRT) in Methods. Within DRT, we obtain an implicit expression of $(Q_{\rm p},Q_{\rm t})$ for an arbitrary protocol $a(t)$ as
\begin{equation} \label{eq:TotalHeatSlow}
\begin{aligned}
	Q_{\rm t} &= (\tau+\tau_{\rm r}) P_0 + \tau_{\rm r} \avg{P}_\text{s}(a_\tau) + B_{\rm t} + \int_0^\tau L(a,\dot{a}) \D t \, ,
	\\
	Q_{\rm p} &= \tau P_0 + B_{\rm t} + \dot a_\tau B_{\rm p} + \int_0^\tau L(a,\dot{a}) \D t \, ,
\end{aligned}
\end{equation}
where $\avg{\cdot}_\text{s}$ denotes a stationary average, and $(B_{\rm t},B_{\rm p})$ are boundary terms set by $(a_0,a_\tau)$ (see Ref.~\cite{SM}, section I). The Lagrangian $L$ is the only contribution depending on the full protocol:
\begin{equation} \label{eq:HeatLagrangian}
	L(a,\dot{a}) \dfn m(a) \dot{a}^2 + \avg{P}_\text{s}(a) \, .
\end{equation}
Within Lagrangian mechanics, $L$ maps into the Lagrangian of a particle with position $a$ and effective mass $m(a)$ subject to the potential $-\avg{P}_\text{s}(a)$. We relate $m(a)$ to some response functions (or, alternatively, some correlation functions~\cite{Davis2024}) measured at fixed $a$ (see Ref.~\cite{SM}, section I).

At large $\tau$, $L$ is dominated by the term $\avg{P}_\text{s}(a)$, so that $(Q_{\rm p},Q_{\rm t})$ scale like $\tau$. At short $\tau$, $L$ is dominated by the term $m(a)\dot a^2$, so that $(Q_{\rm p},Q_{\rm t})$ scale like $1/\tau$; although such a scaling deviates from the expectations for $Q_{\rm p}$ [Fig.~\ref{fig:ProcessHeatVsTotalHeat}], it can still help capture the location of any local minimum. In fact, the crossover between the DRT predictions at large and small $\tau$ always entails a minimum for $(Q_{\rm p}, Q_{\rm t})$ at finite $\tau$. For a given $a(t)$, we deduce from Eqs.~(\ref{eq:TotalHeatSlow}-\ref{eq:HeatLagrangian}) the location of the optimal durations:
\begin{equation}\label{eq:OptimalProcessDuration}
	\tau_{\rm p}^2 = \frac{ \dot a_{s=1} B_{\rm p} + \int_0^1 \dot a^2 m(a) \D s }{ P_0 + \int_0^1 \avg{P}_\text{s}(a) \D s } \, ,
	\quad
	\tau_{\rm t}^2 = \frac{\int_0^1 \dot a^2 m(a) \D s }{ P_0 + \int_0^1 \avg{P}_\text{s}(a) \D s } \, ,
\end{equation}
where we have changed variable $s=t/\tau$, and $\dot a = da/ds$. Therefore, $Q_{\rm p}$ and $Q_{\rm t}$ are not minimum at the same duration $\tau$ in general. Again, let us emphasize that DRT predicts that $Q_{\rm p}$ always features a local minimum: we examine below in detail the conditions under which this prediction can actually fail. On the contrary, $Q_{\rm t}$ always features a minimum and DRT will give good predictions for it close to the quasistatic limit.

In short, the DRT predictions [Eqs.~(\ref{eq:TotalHeatSlow}-\ref{eq:OptimalProcessDuration})] hold for a generic active field theory within our thermodynamically-consistent formulation [Eqs.~(\ref{eq:cons}-\ref{eq:OnsagerMatrix})], yet they rely on assuming that the protocol is close to quasistatic ($\Omega\tau\gg 1$). To deploy DRT on a specific dynamics, the main challenge is to explicitly evaluate the Lagrangian in terms of $a(t)$. In what follows, we demonstrate how DRT can inform the optimization of protocols for a new model of active phase separation.


\subsection*{Field theory of active phase separation}

We now introduce a case study, referred to as Chemical Model B (CMB), which corresponds to a specific active field theory for nonequilibrium phase separation. To this end, we consider the canonical $\phi^4$ free-energy functional~\cite{ChaikinLubensky1995}
\begin{equation}\label{eq:free}
	F([\phi],a) \dfn \int_V \bigg[ \frac{a(t)}{2} \fld^2 + \frac{b}{4} \fld^4 + \frac{\kappa}{2} (\bs\dif\fld)^2 \bigg] \D x \,,
\end{equation}
where $b,\kappa>0$. We focus on the case of one spatial dimension ($d=1$), for which the spurious drift $\bbnu$ [Eq.~\eqref{eq:MatrixDynamics}] vanishes~\cite{Markovich2021}. Here, the parameter $a(t)$ is related to the second virial coefficient, as obtained from coarse-graining procedures~\cite{Grosberg2015, Li2023}, and can be varied in various manners (e.g., controlling particle interactions by changing the pH of tunable colloidal systems \cite{Omar2012}).

To leading order in powers of $\phi$ and its gradients, the simplest form of the coupling term $C$ [Eq.~\eqref{eq:OnsagerMatrix}] reads
\begin{equation}\label{eq:coupling}
	C = \gamma \partial_x \fld \, ,
\end{equation}	
where $\gamma$ is constant. The expression in Eq.~\eqref{eq:coupling} ensures that $C$ vanishes for a constant $\phi$, as expected for homogeneous isotropic systems. The corresponding dynamics of the active field [Eqs.~(\ref{eq:cons}-\ref{eq:MatrixDynamics})] follows as
\begin{equation}\label{eq:cmb}
	\begin{aligned}
	\dot \phi &+ \partial_x J = 0 \,,
	\\
	J &= - \lambda_\phi \partial_x \bigg[ \bigg( a - \frac{\gamma \act}{\mob_\phi} + b \phi^2 - \kappa \partial_x^2 \bigg) \phi \bigg] + T^{1/2} \Lambda_\phi \,,
	\end{aligned}
\end{equation}
where the noise term $\Lambda_\phi$ [Eqs.~(\ref{eq:noise}-\ref{eq:OnsagerMatrix})] is Gaussian with zero mean and correlations given by
\begin{equation}\label{eq:noise_cmb}
	\langle \Lambda_\phi(x,t) \Lambda_\phi(x',t') \rangle = 2 \lambda_\phi \delta(x-x')\delta(t-t') \, .
\end{equation}
Therefore, at the level of the $\phi$-dynamics, the coupling between chemical and active fields [Eq.~\eqref{eq:coupling}] simply amounts to taking $a \to \tilde{a} = a - \gamma \act/\mob_\phi$ in the free energy $F$ [Eq.~\eqref{eq:free}]. Then, the statistics of $\phi$ obeys equilibrium properties (Boltzmann steady-state, fluctuation-dissipation theorem, etc.~\cite{Gardiner}), although the coupled dynamics of $(\phi, n)$ operates away from equilibrium.

\begin{figure}
\includegraphics[width=\linewidth]{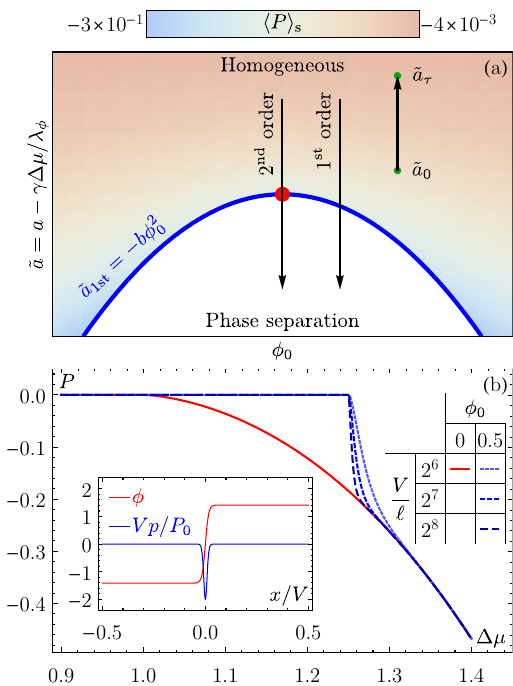}
	\caption{(a)~Phase diagram of Chemical Model B (CMB) [Eqs.~(\ref{eq:free}-\ref{eq:noise_cmb})] in terms of the free-energy parameter $\tilde{a}=a - \gamma \act/\mob_\phi$ and the average density $\fld_0 = \frac 1 V \int_V \phi \D x$. The solid blue line ($\tilde{a}_\text{1st} =  - b \phi_0^2$) separates homogeneous and phase-separated states with a critical point at the red dot. The color map indicates the value of $\avg{P}_\text{s}$ for $T>0$ [Eqs.~(\ref{eq:diss}-\ref{eq:P})].
	(b)~The mean-field ($T=0$) dissipation rate $P$ changes across phase transitions when varying activity $\Delta\mu$, from $P=0$ (homogeneous) to $P<0$ (phase separation). Through the critical point ($\phi_0=0$), $P$ changes smoothly. For a discontinuous transition ($\phi_0=0.5$), $P$ changes more steeply as $V/\ell$ increases. Inset shows, for $\phi_0 = 0$ and in the phase-separated state, the spatial profiles of the density $\phi$ (red solid line) and local power $p$ (blue solid line).
	Parameters: $a = b = \kappa = \mob_\fld = \gamma = 1$, $V/\ell=64$, $\act=3$, (a)~$F(\phi_0,a_0)/T = 50 V/\ell$, (b)~$T = 0$.}
	\label{fig:CMBsteadyState}
\end{figure}

The equilibrium mapping of the active field dynamics allows us to readily deduce the phase diagram~\cite{ChaikinLubensky1995}. For large enough $\Delta\mu$, the system undergoes a separation between high-density ($\phi>0$) and low-density ($\phi<0$) regions [Fig.~\ref{fig:CMBsteadyState}(a)]. Unlike other models of active phase separation~\cite{Cates2015, Nardini2017, Fodor2022}, the activity in CMB does not necessarily stem from microscopic self-propulsion. More generally, activity here results from a coupling of the system with a fuel reservoir; for instance, ATP molecules powering nonequilibrium phase separation in living cells~\cite{Weber2014, Weber2019}.

For a constant $a(t)$, the active work $W_{\rm act}$ quantifies the dissipation required to sustain steady-state activity, and follows from Eqs.~(\ref{eq:ActiveWork}-\ref{eq:ActivePowerDefinition}) as
\begin{equation}\label{eq:diss}
	W_{\rm act} = (\tau+\tau_{\rm r}) (P_0 + \avg{P}_\text{s}) \,,
\end{equation}
where now
\begin{equation}\label{eq:P}
	P = \int_V p \, \D x \,,
	\quad
	p = \frac{\gamma \act}{\mob_\fld} (\partial_x \phi) \left( J - \gamma \act\, \partial_x \phi \right) \, ,
\end{equation}
and $J$ is given in Eq.~\eqref{eq:cmb}. At the mean-field level ($T=0$), the phase-separated profile $\phi(x)$ is known analytically~\cite{ChaikinLubensky1995}, from which we deduce $p(x)$ (see Ref.~\cite{SM}, section IV.A): it vanishes in the bulk phases, and is negative at the interface [Fig.~\ref{fig:CMBsteadyState}(b) inset]. Therefore, the dissipation is maximal in the homogeneous state, where all the fuel energy is wasted in the thermostat, and is reduced for phase separation, where part of the fuel energy serves to shape the density profile [Fig.~\ref{fig:CMBsteadyState}(b)]. In the homogeneous state, our analytical results (see Ref.~\cite{SM}, section IV.B) show that the finite-temperature value of $\avg{P}_\text{s}$ is negative, and its absolute value reduces closer to the phase boundary [Fig.~\ref{fig:CMBsteadyState}(a)], while maintaining $W_\text{act}\geq0$.

The thermodynamically consistent structure of our active field theory [Eqs.~(\ref{eq:cons}-\ref{eq:OnsagerMatrix})] is built to properly account for all sources of dissipation coming from the diffusive and chemical sectors of the dynamics. For CMB [Eqs.~(\ref{eq:free}-\ref{eq:noise_cmb})], the dissipation is locally reduced at interfaces, in line with similar results for another model of active phase separation~\cite{Markovich2021}. As discussed below, the behavior of the steady-state dissipation across the phase diagram determines the shape of the optimal protocol.


\begin{figure}
  \includegraphics[width=\linewidth]{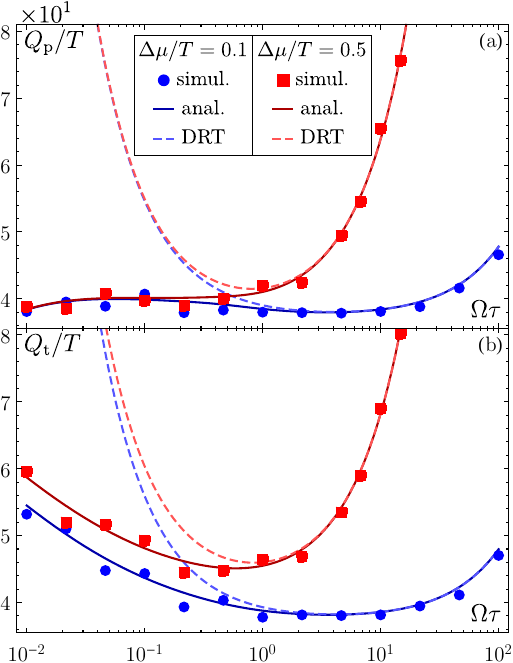}
	\caption{(a)~Protocol heat $Q_{\rm p}$ and (b)~total heat $Q_{\rm t}$ as functions of the protocol duration $\Omega\tau$ for the linear protocol $a_{\rm lin}$ [Eq.~\eqref{eq:naive}] at two levels of activity $\Delta\mu/T$. Markers refer to numerical simulations (code available in~\cite{simul}), solid lines to analytical results, and dashed lines to predictions of DRT [Eqs.~\eqref{eq:TotalHeatSlow}].
	Parameters: $a_0 = 1$, $a_\tau = 10$, $b=\kappa=1$, $\phi_0 = 0.5$, $\mob_\phi = \mob_n = \gamma = 1$, $V/\ell = 64$, $F(\phi_0,a_0)/T = 50 V/\ell$.
	}
	\label{fig:naive}
\end{figure}

\subsection*{Scalings of optimal protocol duration}

For CMB dynamics [Eqs.~(\ref{eq:free}-\ref{eq:noise_cmb})], we consider a protocol that evolves $a(t)$ within the homogeneous state. Without loss of generality, we take $a_\tau>a_0$ [Fig.~\ref{fig:CMBsteadyState}(a)], so that the energy difference $\langle F(a_\tau) - F(a_0)\rangle$ and the heat $(Q_{\rm p},Q_{\rm t})$ are always positive. For simplicity, we first examine the case of the linear protocol:
\begin{equation}\label{eq:naive}
	a_{\rm lin}(t) = a_0 + (t/\tau)(a_\tau-a_0) \, .
\end{equation}
Using numerical simulations (see Ref.~\cite{SM}, section IV.G), we evaluate the heat $(Q_{\rm p},Q_{\rm t})$ [Eqs.~(\ref{eq:TotalHeat}-\ref{eq:work}) and~(\ref{eq:ExternalWork}-\ref{eq:ActivePowerDefinition})] for two levels of activity $\Delta\mu/T=(0.1,0.5)$ [Fig.~\ref{fig:naive}]. In both cases, $Q_{\rm t}$ features a global minimum at an intermediate value of the protocol duration $\tau$. In contrast, $Q_{\rm p}$ has a local minimum for $\Delta\mu/T=0.1$, whereas it increases monotonically for $\Delta\mu/T=0.5$. As previously discussed, these features support the fact that $Q_{\rm t}$ is generally the relevant heat to optimize, since the existence of a minimum is independent of the activity.

To obtain some analytical predictions, we focus on the regime of weak noise where $T$ is much smaller than the local energy scale $(\ell/V)F$, with $\ell$ being the grid spacing, which also corresponds to large system size. Expanding the active field $\phi$ around the mean-field solution $\phi_0 = \frac 1 V \int_V \phi \D x $, the dynamics of each mode $\tilde \phi_k = \frac 1 V \int_V e^{i k x} (\phi-\phi_0)\D x$ is decoupled and linear to leading order in $T$. Our corresponding analytical results for $(Q_{\rm p},Q_{\rm t})$ (see Ref.~\cite{SM}, section III.B) reproduce the curves sampled numerically [Fig.~\ref{fig:naive}]. Alternatively, one can also evaluate $(Q_{\rm p},Q_{\rm t})$ using DRT in the weak-noise regime (see Ref.~\cite{SM}, section III): these predictions are in good agreement with the numerics at large $\Omega\tau$, yet show a discrepancy at small $\Omega\tau$ [Fig.~\ref{fig:naive}], as expected. Moreover, DRT quantitatively captures the locations of $(\tau_{\rm t},\tau_{\rm p})$ only at a small activity ($\Delta\mu/T=0.1$).

\begin{figure*}
	\includegraphics[width=\linewidth]{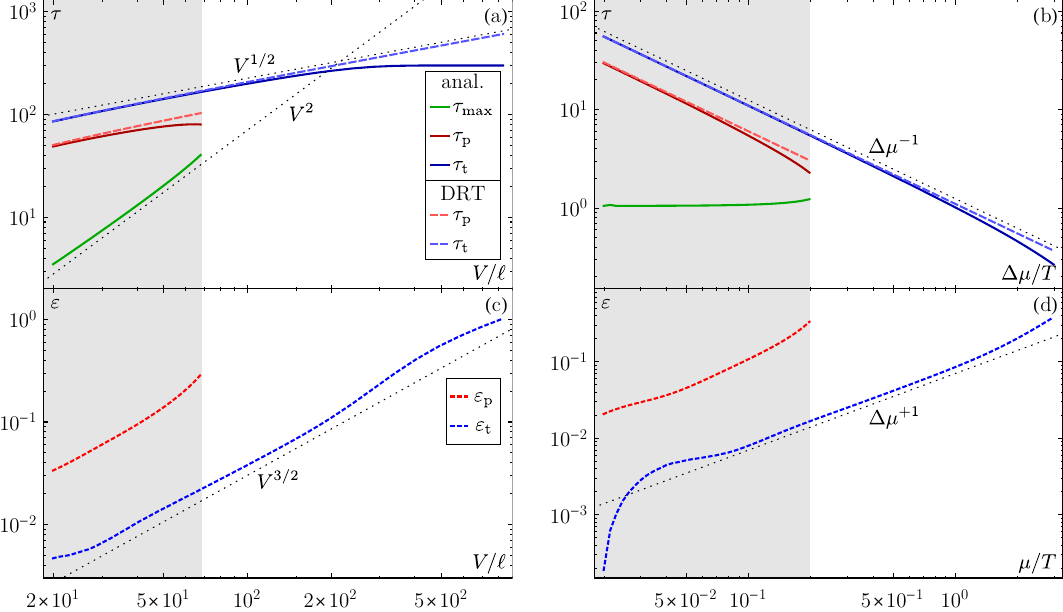}
	\caption{Extremal protocol durations $(\tau_{\rm p},\tau_{\rm t},\tau_{\rm max})$ of the heat $(Q_{\rm p},Q_{\rm t})$ [Fig.~\ref{fig:ProcessHeatVsTotalHeat}] as functions of (a)~system size $V/\ell$ and (b)~activity $\Delta\mu/T$ for the linear protocol $a_{\rm lin}$ [Eq.~\eqref{eq:naive}]. Solid and dashed lines respectively refer to analytical results and DRT predictions [Eq.~\eqref{eq:OptimalProcessDuration}]. Dotted lines are guidelines for scaling behaviors. Beyond the shaded region, $\tau_{\rm p}$ and $\tau_{\rm max}$ coalesce such that $Q_{\rm p}$ becomes monotonic.
    (c-d)~The relative errors $\varepsilon_{\rm x} = \tau_{\rm x}^{\rm DRT} / \tau_{\rm x}^{\rm anal} - 1$, defined for ${\rm x}=({\rm p},{\rm t})$, compare analytical results with DRT predictions taken from (a-b), respectively.
	Parameters: $a_0 = 1$, $a_\tau = 10$, $b = 10$, $\kappa = 1$, $\fld_0 = 0.5$, $\mob_\fld = \mob_n = \gamma = 1$, $F(\phi_0,a_0)/T = 50 V/\ell$,
	(a,c)~$\act/T=0.02$, (b,d)~$V/\ell=10$.
	}
	\label{fig:scalings}
\end{figure*}

We now analyze how the extremal durations $(\tau_{\rm t}, \tau_{\rm p}, \tau_{\rm max})$ [Fig.~\ref{fig:ProcessHeatVsTotalHeat}] scale with activity $\Delta\mu/T$ and system size $V/\ell$. In particular, our aim is to find the regimes where the DRT predictions are accurate. Based on leading-order expansions (see Ref.~\cite{SM}, section IV.D), we obtain the scalings $(B_{\rm p}, m) = {\cal O}(V^2, \Delta\mu^0)$ and $\left(P_0,\avg{P}_\text{s}\right) = {\cal O}(V,\Delta\mu^2)$ at large $V$ and small $\Delta\mu$, from which we deduce the scalings of the minima $(\tau_{\rm p},\tau_{\rm t}) = {\cal O}(V^{1/2},\Delta\mu^{-1})$ using DRT predictions [Eq.~\eqref{eq:OptimalProcessDuration}]. Our analytical results for $(\tau_{\rm p},\tau_{\rm t})$ are in good agreement with DRT for small $\Delta\mu/T$ and moderate $V/\ell$ [Figs.~\ref{fig:scalings}(a-b)], and the relative errors increase with $(\Delta\mu/T, V/\ell)$ [Figs.~\ref{fig:scalings}(c-d)]. Remarkably, the analytical results also suggest the scaling of the maximum $\tau_{\rm max} = {\cal O}(V^2,\Delta\mu^0)$, from which it follows that the local minimum and maximum of $Q_{\rm p}$ [Fig.~\ref{fig:ProcessHeatVsTotalHeat}] always merge at large $(\Delta\mu/T, V/\ell)$ [Figs.~\ref{fig:scalings}(a-b)].

We reveal that DRT becomes less accurate as volume $V$ and activity $\Delta\mu$ increase: this result is consistent with the assumption that DRT breaks down when the relaxation time $\Omega^{-1}$, which decreases with $(V,\Delta\mu)$ (see Ref.~\cite{SM}, section IV.B), becomes comparable with $(\tau_{\rm t}, \tau_{\rm p}, \tau_{\rm max})$. In fact, our DRT scaling predictions are valid for an arbitrary protocol $a(t)$. The scaling $(\tau_{\rm p},\tau_{\rm t})\sim V^{1/2}$ is a consequence of the conservation of particle number: we find a different scaling for active theories of non-conserved fields (see Ref.~\cite{SM}, section V). In contrast, the scaling $(\tau_{\rm p},\tau_{\rm t})\sim\Delta\mu^{-1}$ is robust: it embodies the competition between internal activity and external perturbation, respectively regulating the small- and large-$\tau$ regimes, inherent to any active system~\cite{Davis2024}.

\subsection*{Crossover between optimal protocols}

\begin{figure}
	\includegraphics[width=\linewidth]{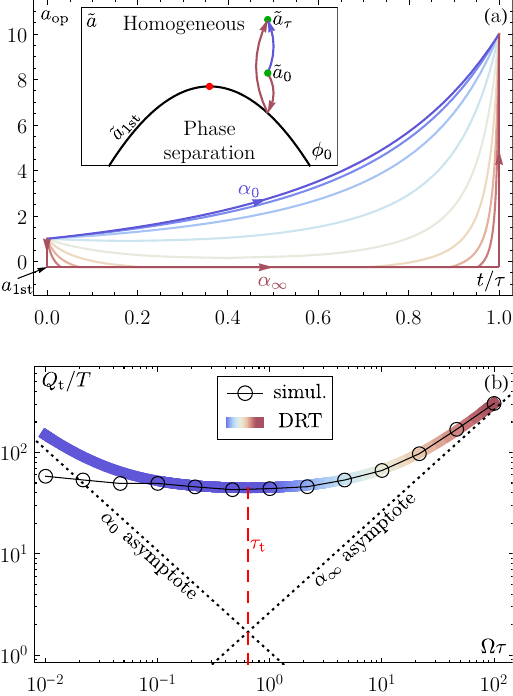}
	\caption{(a)~Optimal protocol $a_\text{op}$, minimizing $Q_{\rm t}$ for a given protocol duration $\tau$, as a function of time $t/\tau$. Colors refer to values of $\tau$, from $\Omega\tau\ll 1$ (dark blue, master curve $\alpha_0$) to $\Omega\tau\gg 1$ (dark red, master curve $\alpha_\infty$). Inset shows the protocols $(\alpha_0, \alpha_\infty)$ in the phase diagram [Fig.~\ref{fig:CMBsteadyState}].
		(b)~Total heat $Q_{\rm t}$, optimized for a given $\tau$, as a function of protocol duration $\Omega\tau$. The solid line, with color code corresponding to protocols in (a), and markers respectively represent DRT predictions and numerical simulations. The dashed black lines are the asymptotic behaviors extracted from master curves in (a). The red horizontal dashed line highlights the (global) optimal protocol duration $\tau_{\rm t}$ [Eq.~\eqref{eq:tau_approx}].		
		 Same parameters as in Fig.~\ref{fig:naive}.
	}
	\label{fig:optimal}
\end{figure}

We continue by examining the DRT predictions for the optimal protocol $a_\text{op}(t)$ minimizing $(Q_{\rm p},Q_{\rm t})$ [Eq.~\eqref{eq:TotalHeatSlow}] at a given duration $\tau$. This amounts to optimizing the Lagrangian $L$ [Eq.~\eqref{eq:HeatLagrangian}], since the rest of the terms in $Q_{\rm t}$ [Eq.~\eqref{eq:TotalHeatSlow}] are independent of the protocol shape $a(t)$. The optimization of $L$ is obtained explicitly by solving the corresponding Euler-Lagrange equation (see Ref.~\cite{SM}, section IV.E). Since $Q_{\rm p}$ and $Q_{\rm t}$ [Eq.~\eqref{eq:TotalHeatSlow}] have the same functional dependence on $a(t)$, we attribute them the same optimal protocol $a_{\rm op}(t)$. Again, we focus here on the CMB dynamics in the weak-noise regime [Eqs.~(\ref{eq:free}-\ref{eq:noise_cmb})] with $a(t)$ evolving within the homogeneous state, see Fig.~\ref{fig:optimal}(a).

In general, the shape of the optimal protocol changes with $\tau$, that is $a_{\rm op} = a_{\rm op}(t/\tau,\tau)$. Interestingly, we observe a crossover between two types of optimal protocols: at small $\tau$, $a_\text{op}$ tends to the master curve $\alpha_0(t/\tau)$, which monotonically increases from $a_0$ to $a_\tau$; at large $\tau$, $a_\text{op}$ tends to the master curve $\alpha_\infty(t/\tau)$, which has a non-monotonic behavior [Fig.~\ref{fig:optimal}(a)]. We rationalize this crossover by analyzing how $L$ scales with $\tau$. Given that the optimal protocol takes the form $a_{\rm op}(t) = \alpha(t/\tau)$ at large or small $\tau$, the corresponding contribution to $(Q_{\rm p},Q_{\rm t})$ can be written as
\begin{equation}\label{eq:scaling}
	\int_0^\tau L(a_{\rm op}, \dot a_{\rm op}) \D t = \int_0^1 \bigg[ \frac{m(\alpha) \dot \alpha^2}{\tau} + \tau \avg{P}_\text{s}(\alpha) \bigg] \D s \, .
\end{equation}
The first term in Eq.~\eqref{eq:scaling} dominates at small $\tau$: it has the same functional form as for a passive system~\cite{Sivak2012}, and accordingly yields the monotonic protocol $\alpha_0(t/\tau)$. The second term in Eq.~\eqref{eq:scaling} dominates at large $\tau$: it is optimal when the protocol minimizes the steady-state power $\avg{P}_\text{s}$ [Eq.~\eqref{eq:P}]. For CMB, $\avg{P}_\text{s}$ is minimal at the phase boundary [Fig.~\ref{fig:CMBsteadyState}(a)]. Therefore, the corresponding optimal protocol consists of setting $a(t)$ close to this boundary for as long as possible, leading to the master curve $\alpha_\infty(t/\tau)$ [Fig.~\ref{fig:optimal}(a)].

The predictions for the optimal heat $Q_{\rm t}$, which minimize heat for all protocol shapes and durations, follow by substituting $a_{\rm op}$ into Eq.~\eqref{eq:TotalHeatSlow}. The agreement with numerical measurements breaks down at small $\Omega\tau$, as expected, yet the discrepancy remains negligible close to the (global) optimal duration $\tau_{\rm t}$ [Fig.~\ref{fig:optimal}(b)]. By matching the asymptotic behaviors of $Q_{\rm t}$ at large and small $\tau$, we approximate $\tau_{\rm t}$ [Eq.~\eqref{eq:OptimalProcessDuration}] as
\begin{equation}\label{eq:tau_approx}
	\tau_{\rm t}^2 \approx \frac{ \int_0^1 \dot \alpha_0^2 m(\alpha_0) \D s }{ P_0 + \int_0^1 \avg{P}_\text{s}(\alpha_\infty) \D s} \, .
\end{equation}
In practice, Eq.~\eqref{eq:tau_approx} reproduces the value estimated numerically (see Ref.~\cite{SM}, section IV.F for more details) for our specific parameters [Fig.~\ref{fig:optimal}(b)]: DRT predictions quantitatively capture the global minimum of $Q_{\rm t}$ whenever their regime of validity contains $\tau_{\rm t}$.

Therefore, we conclude that the crossover between optimal protocols is governed by the competition between the dissipation stemming from either external driving or internal activity, respectively dominant at small and large $\tau$. This crossover is expected to appear generically when controlling homogeneous states in a broad class of active systems, beyond the specific case of CMB (see Ref.~\cite{SM}, section IV.E).

\section*{Discussion}

In this work, we have examined how to quantify and optimize the heat dissipated by complex active systems under external perturbation. Specifically, for thermodynamically consistent field theories of active matter, we have distinguished two definitions of heat. The protocol heat $Q_{\rm p}$ is dissipated only during the protocol, and may feature a {\em local} minimum as a function of the protocol duration $\tau$ only under specific conditions. In contrast, the total heat $Q_{\rm t}$ is dissipated during both the protocol and its post-relaxation, and systematically features a {\rm global} minimum. Therefore, our study goes beyond previous works which mostly focused on $Q_{\rm p}$~\cite{Blaber2023, Davis2024}: we have here highlighted the essential role of post-relaxation in quantifying the energetics of active systems.

We have put forward a systematic framework, inspired by previous works~\cite{Sivak2012, Davis2024}, to optimize the external perturbation using dynamical response theory. Other frameworks unrealistically assume that the operator has full control over the system (namely, by completely changing the rules of the dynamics)~\cite{Benamou2000, Aurell2011}; this assumption is particularly useful to derive bounds on dissipation~\cite{Dechant2020, Saito2023, Rotskoff2023}. In contrast, our predictions apply here to the manipulation of some experimentally accessible parameters (namely, by changing the shape of a given free-energy). Remarkably, our framework allows one to evaluate: (i) the optimal duration of operation for a given protocol, (ii) the optimal protocol for a fixed duration, and (iii) the overall duration that gives minimal heat and the corresponding optimal protocol.

For the dynamics of a conserved scalar field that exhibits a nonequilibrium phase-separation, our scaling analysis predicts that $Q_{\rm p}$ does not feature any minimum at large activity $\Delta\mu$ and volume $V$. A similar analysis for non-conserved scalar fields yields a different scaling for $V$ (see Ref.~\cite{SM}, section V), so that $Q_{\rm p}$ may show a local minimum even at large $V$. These results pave the way towards a systematic scaling analysis to inform protocol optimization in a broader class of active field theories. For instance, our approach could be deployed in polar active theories of the flocking transition~\cite{Chate2020}, for which a thermodynamically-consistent description has been recently proposed~\cite{Agranov2024}.

When driving homogeneous active systems, we have revealed a crossover between two families of optimal protocols. In the specific case of a conserved scalar field, we have shown that this crossover embodies the competition between the internal and external contributions to dissipation, stemming respectively from activity and perturbation. Remarkably, since this competition is inherent to a broad class of active systems~\cite{Davis2024}, optimizing protocols between homogeneous states should generically yields a similar crossover. To go beyond our results, it remains to address protocols with  multiple control parameters, for instance to build thermodynamic cycles~\cite{Pietzonka2018b, Frim2022a, Frim2022b}. Moreover, it would be interesting to examine how our framework can be extended to protocols crossing phase transitions.

\acknowledgments{AS and TM acknowledge funding from the Israel Science Foundation (Grant No. 1356/22). ET acknowledges funding from EPSRC (Grant No. EP/W027194/1). EF acknowledges funding from the Luxembourg National Research Fund (FNR), grant reference 14389168. This research was supported in part by grant no. NSF PHY-2309135 to the Kavli Institute for Theoretical Physics (KITP).
}


\section*{Methods}

\subsection*{Stochastic thermodynamics}

Stochastic thermodynamics provides a way to calculate the average heat dissipated into the environment during the time $\tau+\tau_{\rm r}$ as~\cite{Seifert2012, Cates2022}:
\begin{equation}\label{eq:HeatDefinition}
	Q_{\rm t} \dfn T \Avg{ \ln \frac{\mathcal P_+(\tau+\tau_{\rm r})}{\mathcal P_-(\tau+\tau_{\rm r})} } \, ,
\end{equation}
where $\mathcal P_+$ and $\mathcal P_-$ are the probabilities of observing a given trajectory with the forward and the time-reversed dynamics, respectively. These probabilities take the form $\mathcal P_{\pm}(t) \sim e^{- A_{\pm}(t)}$ in terms of the dynamical action:
\begin{equation}\label{eq:ForwardDynamicalAction}
	A_\pm(t) = \frac{1}{4T} \int_0^t \int_V \left(\bb L \cdot \bb\frc \mp \bb\crt \right) \cdot \bb L^{-1} \cdot \left( \bb L \cdot \bb\frc \mp \bb\crt \right) \D{\bf r} \D t' \, ,
\end{equation}
where $\bb L^{-1}$ is the inverse of the Onsager matrix $\bb L$. The expression in Eq.~\eqref{eq:ForwardDynamicalAction} is given in the Stratonovitch convention~\cite{Lau2007,Sorkin2025}, where we have omitted some terms which are invariant under time-reversal. This connection between path probabilities and the average heat dissipation is quite general and extends even to systems in which the Einstein relation does not hold~\cite{Sorkin2024}.

Substituting Eq.~\eqref{eq:ForwardDynamicalAction} into Eq.~\eqref{eq:HeatDefinition} yields the total heat $Q_{\rm t}$ in terms of the flux $\bb\crt$ and force $\bb\frc$ [Eq.~\eqref{eq:heat}], which can also be written as
\begin{equation} \label{eq:HeatFromFluxes&Forces}
	Q_{\rm t} = \Avg{ \int_0^{\tau+\tau_{\rm r}} \int_V \bigg( \act \, \dot \fue -  \frac{\delta F}{\delta\fld} \dot\fld \bigg) \D{\bf r} \D t  } \, ,
\end{equation}
where we have used the definitions $\bb\crt = (\bs\crt,\dot\fue)$ and $\bb\frc =(-\bs\dif (\delta F/\delta\fld), \act)$. A related expression has been derived for the heat in the absence of perturbation ($\dot a=0$) in terms of steady-state  averages~\cite{Markovich2021}. For a given protocol $a(t)$, the standard chain rule, valid within the Stratonovitch convention~\cite{Gardiner}, yields
\begin{equation} \label{eq:FreeEnergyIdentity}
	\int_V \frac{\delta F}{\delta\fld} \dot\fld \, \D{\bf r} = \dot F - \dot a \pd_a F \, ,
\end{equation}
from which we deduce
\begin{equation}\label{eq:FirstLawPlusSpuriousHeat}
	Q_{\rm t} = W_\text{ext} + W_\text{act} + E(t=0) - E(t=\tau+\tau_{\rm r}) \, ,
\end{equation}
where we have used $E = \langle F\rangle$, along with the definitions of the external work $W_\text{ext}$ [Eq.~\eqref{eq:ExternalWork}] and the active work $W_\text{act}$ [Eq.~\eqref{eq:ActiveWork_0}]. Substituting the dynamics [Eqs.~(\ref{eq:MatrixDynamics}-\ref{eq:OnsagerMatrix})] into the definition of $W_\text{act}$, we obtain
\begin{equation} \label{eq:ActiveWorkWithSpuriousPower}
	W_{\rm act} = (\tau+\tau_{\rm r}) P_0 + \Avg{ \int_0^{\tau+\tau_{\rm r}} (P+P_\text{spu}) \D t } \, ,
\end{equation}
in terms of
\begin{equation}\label{eq:spuriousActivePower}
	\begin{aligned}
		P_\text{spu} &\dfn T^{1/2} \frac{\act}{\mob_\fld} \int_V ( \mob_\fld \noi_\fue - \bs\cpl \cdot \bs\noi_\fld ) \D{\bf r}
		\\
		&\quad + T \frac{\act}{\mob_\fld} \int_V ( \mob_\fld \spu_\fue - \bs\cpl \cdot \bs\spu_\fld ) \D{\bf r} \, .
	\end{aligned}
\end{equation} 
The stochastic integrals in Eq.~\eqref{eq:ActiveWorkWithSpuriousPower} are interpreted within the Stratonovitch convention. Consequently, we demonstrate in Ref.~\cite{SM}, section VI that $\avg{\int P_\text{spu} \D t}=0$, so Eq.~\eqref{eq:ActiveWork} indeed corresponds to $W_{\rm act}$.
In fact, the spurious drift $\bbnu$ ensures thermodynamic consistency of the dynamics [Eq.~\eqref{eq:MatrixDynamics}], but the active work [Eq.~\eqref{eq:ActiveWork}] does not explicitly depend on $\bbnu$, as expected for noise-averaged quantities.


\subsection*{Dynamical response theory}

Evaluating the expressions for heat [Eq.~\eqref{eq:heat}] and work [Eqs.~(\ref{eq:ExternalWork}-\ref{eq:ActiveWork})] requires computing the non-steady-state average $\Avg{O}(t)$ of some observables $O$ at an arbitrary time $t$. In general, $\Avg{O}(t)$ depends on the entire history of the protocol $a(t')$ for $t'\in[0,t]$. Inspired by previous studies~\cite{Sivak2012, Davis2024}, in the regime of slow protocols ($\Omega\tau\ll1$), we demonstrate how to express $\Avg{O}(t)$ in terms of some specific response functions, or equivalently in terms of unperturbed correlation functions. Importantly, our approach assumes that the change in $a$ is slow compared with the system relaxation time $\Omega^{-1}$, yet the overall change $a(\tau)-a(0)$ can be large.

Let us split the protocol $a(t) = \sum_n \Delta a(t_n,t_{n-1})$ as a series of discrete jumps $\Delta a(t_n,t_{n-1}) = a(t_n) - a(t_{n-1})$ with durations $\delta t\ll\tau$, so that $t_n=n\delta t$~\cite{Bonanca2014}. We approximate $\Avg{O}(t_n)$ close to $\Avg{O}(t_{n-1})$ as~\cite{Kubo1991}
\begin{equation}\label{eq:DRTdiscretized}
\begin{aligned}
	\Avg{O}(t_{n}) &\approx \Avg{O}(t_{n-1})
	\\
	&\quad + \int^{t_n}_{t_{n-1}} \D t' \, \Delta a(t',t_n) \bigg[ R_O^{(1)}(t_n-t')  
	\\
	&\quad + \int^{t_n}_{t_{n-1}} \D t'' \, \Delta a(t',t_n) R_O^{(2)}(t_n-t',t_n-t'') \bigg] \, ,
\end{aligned}
\end{equation}
where $R_O^{(1)}$ and $R_O^{(2)}$ are the first- and second-order response functions, given respectively by
\begin{subequations}
\begin{align}
R^{(1)}_O(a(t);t-t_1) &\dfn \left. \frac{\delta \Avg{O(t)}}{\delta a(t_1)} \right|_{a(t_1) \to a(t)}, \\
R^{(2)}_O(a(t);t-t_1,t-t_2) &\dfn \left. \frac{\delta^2 \Avg{O(t)}}{\delta a(t_2) \delta a(t_1)} \right|_{\substack{a(t_1) \to a(t) \\ a(t_2) \to a(t)}}.
\end{align}    
\end{subequations}
Provided that the system relaxes faster than the rate of change of $a$ within each timestep ($\Omega\delta t \gg 1$), we approximate $\langle O\rangle(t_{n-1}) \approx \langle O\rangle_{\rm ss}(a(t_{n-1}))$ as a steady-state average, and we extend the lower limits of integration in Eq.~\eqref{eq:DRTdiscretized} as $t_{n-1}\to-\infty$. Additionally, assuming that the protocol $a(t)$ is smooth, we expand
\begin{equation}\label{eq:exp}
	\Delta a(t',t_n) \approx ( t'-t_n ) \dot a(t_{n}) + ( t'-t_n )^2 \ddot a(t_{n})/2 \, .
\end{equation}
Substituting this expansion in the response framework [Eq.~\eqref{eq:DRTdiscretized}], we deduce
\begin{equation}\label{eq:DRTcontinuous}
\begin{aligned}
	\Avg{O}(t) &\approx \Avg{O}_\text{s}\bigl(a(t)\bigr) + \dot a(t) \zeta_O^{(1,1)}\bigl(a(t)\bigr)
	\\
	&\quad + \ddot a(t) \zeta_O^{(2,1)}\bigl(a(t)\bigr) + \dot a^2(t) \zeta_O^{(2,2)}\bigl(a(t)\bigr) \, ,
\end{aligned}
\end{equation}
where we have taken $t\equiv t_n \approx t_{n-1}$, and introduced 
\begin{equation}
\begin{aligned}
	\zeta_O^{(1,1)} &= - \int_0^\infty t' R_O^{(1)}(t') \D t' \, ,
	\quad
	\zeta_O^{(2,1)} = \int_0^\infty \frac{t'^2}{2} R_O^{(1)}(t') \D t' \, ,
	\\
	\zeta_O^{(2,2)} &= \iint_0^\infty \frac{t' t''}{2} R_O^{(2)}(t',t'') \D t' \D t'' \, .
\end{aligned}
\end{equation}
Note that the observable-specific quantities in the rhs of Eq.~\eqref{eq:DRTcontinuous}, namely $\Avg{O}_\text{s}$ and $\zeta_O^{(m,n)}$, depend only on $a(t)$ and not on its derivatives $(\dot a, \ddot a)$. Finally, substituting Eq.~\eqref{eq:DRTcontinuous} into the expression of heat [Eq.~\eqref{eq:heat}] yields the decomposition given in Eq.~\eqref{eq:TotalHeatSlow}; see Ref.~\cite{SM}, section I for details.


\bibliography{bib_v12.bib}

\end{document}


\title{Supplemental Material for ``Control of active field theories at minimal dissipation''}

\author{Artur Soriani}
\affiliation{School of Mechanical Engineering, Tel Aviv University, Tel Aviv 69978, Israel}
\affiliation{Center for Physics and Chemistry of Living Systems, Tel Aviv University, Tel Aviv 69978, Israel}

\author{Elsen Tjhung}
\affiliation{School of Mathematics and Statistics, The Open University, Walton Hall, Milton Keynes, MK7 6AA, United Kingdom}

\author{Étienne Fodor}
\affiliation{Department of Physics and Materials Science, University of Luxembourg, L-1511 Luxembourg City, Luxembourg}

\author{Tomer Markovich}
\affiliation{School of Mechanical Engineering, Tel Aviv University, Tel Aviv 69978, Israel}
\affiliation{Center for Physics and Chemistry of Living Systems, Tel Aviv University, Tel Aviv 69978, Israel}

\date{\today}

\onecolumngrid

\maketitle

\tableofcontents

\section{Protocol and total heat using dynamic response theory} \label{sec:DRTheats}

In this section, we use Dynamic Response Theory (DRT) to derive Eq.~(12) of the main text, which contain the protocol and total heat for slow processes.
Our starting point is Eq.~(1) of the main text, the first law, where we can use $E = \avg{F}$ and the expressions for the work (Eq.~(2)), subdivided into external work $W_\text{ext}$ (Eq.~(8)) and active work $W_\text{act}$ (Eq.~(10)), to find expressions for the protocol heat $Q_\text{p}$ and the total heat $Q_\text{t}$.
This leads to
\begin{subequations} \label{eq:heats}
\begin{align}
\label{eq:protocolHeat}
Q_\text{p} & = \tau P_0 + \int_0^\tau \avg{P}(t) \, \D t + \int_0^\tau \dot a(t) \avg{\pd_a F}(t) \, \D t - \big( \avg{F}(\tau) - \avg{F}_\text{s}(a_0) \big) \\
\shortintertext{and}
\label{eq:totalHeat}
Q_\text{t} & = (\tau + \tau_\text{r}) P_0 + \int_0^{\tau} \avg{P}(t) \, \D t + \int_{\tau}^{\tau + \tau_\text{r}} \avg{P}(t) \, \D t + \int_0^\tau \dot a(t) \avg{\pd_a F}(t) \, \D t - \big( \avg{F}_\text{s}(a_\tau) - \avg{F}_\text{s}(a_0) \big),
\end{align}
\end{subequations}
where $P_0 = \mob_n V \act^2$, $\avg{\cdot}_\text{s}$ denotes steady-state averages, while all the other averages are non-stationary.
In Eq.~\eqref{eq:totalHeat}, the integral of $\avg{P}$ was split into two because its expressions in the manipulation process ($0 \leq t \leq \tau$) and in the relaxation process ($\tau \leq t \leq \tau+\tau_\text{r}$) are fundamentally different.

Let us first calculate the averages during the manipulation process.
This is where DRT applies, and we can use Eq.~(34) of the main text for the three relevant observables, that is,
\begin{subequations} \label{eq:DRT_continuous}
\begin{align}
\avg{F}(t) &= \avg{F}_\text{s}(a(t)) + \dot a(t) \zeta_F^{(1,1)}(a(t)) + \mathcal O(\ddot a, \dot a^2), \\
\avg{\pd_a F}(t) &= \avg{\pd_a F}_\text{s}(a(t)) + \dot a(t) \zeta_{\pd_a F}^{(1,1)}(a(t)) + \mathcal O(\ddot a, \dot a^2), \\
\shortintertext{and}
\avg{P}(t) &= \avg{P}_\text{s}(a(t)) + \dot a(t) \zeta_P^{(1,1)}(a(t)) + \ddot a(t) \zeta_P^{(2,1)}(a(t)) + \dot a^2(t) \zeta_P^{(2,2)}(a(t)) + \mathcal O(\dddot a, \ddot a \dot a, \dot a^3).
\end{align}
\end{subequations}
Here, we kept the terms necessary to make Eqs.~\eqref{eq:heats} correct up to $\mathcal O(\tau^{-2})$, assuming $\dot a \sim \tau^{-1}$ and $\ddot a \sim \tau^{-2}$.
The expressions for the integrated response functions $\zeta_O^{(m,n)}$ are given in Eq.~(35) of the main text.

To make Eqs.~\eqref{eq:DRT_continuous} valid for $t=0$, we need to enforce $\zeta_O^{(m,n)}(a_0) \dfn 0$ for any $m$ and $n$ (because we assumed $\avg{O}(t_0) = \avg{O}_\text{s}(a_0)$ from the beginning).
Once this is done, placing Eqs.~\eqref{eq:DRT_continuous} into Eqs.~\eqref{eq:heats} and rearranging leads us to
\begin{subequations} \label{eq:DRTheats}
\begin{align}
\label{eq:DRTprotocolHeat}
Q_\text{p} & = \tau P_0 + B_\text{t} + \dot a(\tau) B_\text{p} + \int_0^\tau L\big( a(t),\dot a(t) \big) \, \D t \\
\shortintertext{and}
\label{eq:DRTtotalHeat0}
Q_\text{t} & = (\tau + \tau_\text{r}) P_0 + B_\text{t} + \dot a(\tau) \zeta_P^{(2,1)}(a_\tau) + \int_{\tau}^{\tau + \tau_\text{r}} \avg{P}(t) \, \D t + \int_0^\tau L\big( a(t),\dot a(t) \big) \, \D t,
\end{align}
\end{subequations}
where we defined
\begin{subequations} \label{eq:heatFactors}
\begin{align}
B_\text{t} & \equiv \int_{a_0}^{a_\tau} \left( \avg{\pd_a F}_\text{s}(a) + \zeta_P^{(1,1)}(a) \right) \, da - \Big( \avg{F}_\text{s}(a_\tau) - \avg{F}_\text{s}(a_0) \Big), \\
B_\text{p} & \equiv \zeta_P^{(2,1)}(a_\tau) - \zeta_F^{(1,1)}(a_\tau), \\
\label{eq:DRTlagrangian}
L(a,\dot a) & \equiv m(a) \dot a^2 + \avg{P}_\text{s}(a), \\
\shortintertext{and}
m(a) & \equiv \zeta_{\pd_a F}^{(1,1)}(a) + \zeta_P^{(2,2)}(a) - \pd_a \zeta_P^{(2,1)}(a).
\end{align}
\end{subequations}

Equation~\eqref{eq:DRTprotocolHeat} is already in the form shown in Eqs.~(12) of the main text.
However, Eq.~\eqref{eq:DRTtotalHeat0} still needs the calculation of the integral of $\avg{P}$ in the relaxation process, which we turn to now.
In Sec.~\ref{sec:posthumousHeat} we prove, using the relaxation dynamics of the system, that
\begin{align} \label{eq:activePowerRelaxationIntegral}
\dot a(\tau)\zeta^{(2,1)}_P(a_\tau) + \int_{\tau}^{\tau+\tau_\text{r}} \Avg{ P(t) } \D t = \tau_\text{r} \Avg{ P }_\text{s}(a_\tau),
\end{align}
Finally, placing Eq.~\eqref{eq:activePowerRelaxationIntegral} into Eq.~\eqref{eq:DRTtotalHeat0}, we reach
\begin{align} \label{eq:DRTtotalHeat}
Q_\text{t} & = (\tau + \tau_\text{r}) P_0 + \tau_\text{r} \Avg{ P }_\text{s}(a_\tau) + B_\text{t} + \int_0^\tau L\big( a(t),\dot a(t) \big) \, \D t,
\end{align}
which is the expression for the total heat we see in Eq.~(12) of the main text.

\section{Response functions \label{sec:ResponseFunctions}}

The goal in this section is to obtain practical expressions for the response functions, given our active field dynamics.
From Eqs.~(32) of the main text, the formal definitions of the first and second-order response functions of a generic observable $O$ are
\begin{subequations} \label{eq:responseFunctions}
\begin{align}
\label{eq:1stReponseFunction}
R^{(1)}_O(a(t);t-t_1) & \dfn \left. \frac{\delta \Avg{O(t)}}{\delta a(t_1)} \right|_{a(t_1) \to a(t)}, \\
\shortintertext{and}
\label{eq:2ndResponseFunction}
R^{(2)}_O(a(t);t-t_1,t-t_2) & \dfn \left. \frac{\delta^2 \Avg{O(t)}}{\delta a(t_2) \delta a(t_1)} \right|_{\substack{a(t_1) \to a(t) \\ a(t_2) \to a(t)}},
\end{align}
\end{subequations}
where $t_1$ and $t_2$ are dummy variables, with no relation to the discretized times $t_n$.
To avoid any confusions, we only consider situations in which the time arguments are in decreasing order, i.e., $R^{(1)}_O(a;t-t_1)$ for $t \geq t_1$ and $R^{(2)}_O(a;t-t_1,t-t_2)$ for $t \geq t_1 \geq t_2$.

These definitions can be used to arrive at more practical expressions for the response functions \cite{Davis2024}.
We start from the path integral definition of the average over noise realizations,
\begin{align*}
\Avg{O(t)} & \dfn \int O(t) N^{-1}(t) e^{- A_+[\bb\crt](t)} \mathcal D\noi, \\
N(t) & \dfn \int e^{- A_+[\bb\crt](t)} \mathcal D\noi,
\end{align*}
where $A_+$ is the forward dynamical action given in Eq.~(25) of the main text.
The dependence on $a$ lies only on $N$ and $e^{- A_+}$, so that Eq.~\eqref{eq:1stReponseFunction} gives
\begin{align}
\notag
R^{(1)}_O(a(t);t-t_1) & = \int O(t) \left[ \frac{\delta N^{-1}(t)}{\delta a(t_1)} e^{- A_+[\bb\crt](t)} + N^{-1}(t) \frac{\delta e^{- A_+[\bb\crt](t)}}{\delta a(t_1)} \right]_{a(t_1) \to a(t)} \mathcal D\noi \\
\notag
& = \int O(t) \left[ - N^{-2}(t) \frac{\delta N(t)}{\delta a(t_1)} e^{- A_+[\bb\crt](t)} - N^{-1}(t) e^{- A_+[\bb\crt](t)} \frac{\delta A_+[\bb\crt](t)}{\delta a(t_1)} \right]_{a(t_1) \to a(t)} \mathcal D\noi \\
\notag
& = \int \left[ O(t) N^{-1}(t) \Avg{ \frac{\delta A_+[\bb\crt](t)}{\delta a(t_1)} } e^{- A_+[\bb\crt](t)} - O(t) \frac{\delta A_+[\bb\crt](t)}{\delta a(t_1)} N^{-1}(t) e^{- A_+[\bb\crt](t)} \right]_{a(t_1) \to a(t)} \mathcal D\noi \\
\label{eq:1stResponseFunctionFromGeneralizedForce}
& = \cor{ O(t) G(t_1) }_{a(t_1) \to a(t)},
\end{align}
where we used
\[
\frac{\delta N(t)}{\delta a(t_1)} = - N(t) \Avg{ \frac{\delta A_+[\bb\crt](t)}{\delta a(t_1)} },
\]
from the second to the third line, and we defined the generalized force (omitting functional dependencies)
\begin{equation} \label{eq:GeneralizedForceDefinition}
G(t_1) \dfn - \frac{\delta A_+}{\delta a(t_1)}
\end{equation}
and the two-entry correlation
\begin{equation} \label{eq:TwoEntryCorrelation}
\cor{ O_1 O_2 } \dfn \Avg{O_1 O_2} - \Avg{O_1} \Avg{O_2}.
\end{equation}
Similarly, for the second-order response function, Eq.~\eqref{eq:2ndResponseFunction} gives
\begin{equation} \label{eq:2ndResponseFunctionFromGeneralizedForce}
R^{(2)}_O(a(t);t-t_1,t-t_2) = \cor{ O(t) \frac{\delta G(t_1)}{\delta a(t_2)} }_{\substack{a(t_1) \to a(t) \\ a(t_2) \to a(t)}}
+ \cor{ O(t) G(t_1) G(t_2) }_{\substack{a(t_1) \to a(t) \\ a(t_2) \to a(t)}},
\end{equation}
with the three-entry correlation defined as
\begin{equation} \label{eq:ThreeEntryCorrelation}
\cor{ O_1 O_2 O_3 } \dfn \Avg{O_1 O_2 O_3} + 2 \Avg{O_1} \Avg{O_2} \Avg{O_3}
- \Avg{O_1} \Avg{O_2 O_3} - \Avg{ O_2 } \Avg{O_3 O_1} -  \Avg{O_3} \Avg{O_1 O_2}.
\end{equation}

Looking at Eq.~(31) of the main text, we see that making $a(t_1) \to a(t)$ is the same as taking a steady-state average.
This allows us to write the response function of Eq~\eqref{eq:1stResponseFunctionFromGeneralizedForce} as a steady-state correlation function, i.e.,
\begin{equation} \label{eq:CorrelationFunction}
R^{(1)}_O(a;t-t_1) = \cor{ O(t) G(t_1) }_\text{s}(a),
\end{equation}
and the same is valid for higher orders.
Furthermore, Eq.~(25) of the main text shows that the action is a functional of $a(t_1)$ (through $\bb\frc$), but the integrand does not depend on time derivatives of $a(t_1)$, which simplifies the functional derivative as simply the partial derivative of the integrand.
Combining Eqs.~(25) of the main text and~\eqref{eq:GeneralizedForceDefinition}, we get
\begin{equation} \label{eq:FunctionalToPartialDerivative}
G(t) = - \frac{1}{4T} \frac{\pd}{\pd a} \int_V \left\{ \left[ \bb\crt^\trp - \bb\frc^\trp \cdot \bb L^\trp \right]
\cdot \left[ \bb L^{-1} \cdot \bb\crt - \bb\frc \right] + \mathcal{O}(T) \right\}\dpos,
\end{equation}
where $\mathcal{O}(T)$ stands for contributions that vanish at $T=0$. Here, we refrain from writing these terms explicitly, because they are irrelevant in the following small noise treatment (see Sec.~\ref{sec:smallNoiseExpansion}).

Equation~\eqref{eq:FunctionalToPartialDerivative} shows a product of two binomials and thus four terms, only three of which depend explicitly on $a$ (through $\bb F$).
In two of these three terms, the Onsager matrix completely disappears (because $\bb L^{\trp} = \bb L$), and thus the dynamics of matter and fuel are uncoupled.
Since the fuel dynamics only depends on $a$ through the coupling to active system, its contribution vanishes in these two terms.
The only remaining term is $\bb\frc^\trp \cdot \bb L \cdot \bb\frc$, whose derivative is
\begin{equation} \label{eq:ForcesPartialDerivative}
\frac{\pd}{\pd a} \left[ \bb\frc^\trp \cdot \bb L \cdot \bb\frc \right] = \frac{\pd \bb\frc^\trp}{\pd a} \cdot \bb L \cdot \bb\frc + \bb\frc^\trp \cdot \bb L \cdot \frac{\pd \bb\frc}{\pd a}.
\end{equation}
Take the first term in the RHS of Eq.~\eqref{eq:ForcesPartialDerivative}.
In explicit matrix notation, it reads
\begin{align*}
\frac{\pd \bb\frc^\trp}{\pd a} \cdot \bb L \cdot \bb\frc
& = \begin{pmatrix}
\frac{\pd }{\pd a} \dif \frac{\delta F}{\delta\fld}, & 0
\end{pmatrix}
\cdot
\begin{pmatrix}
\mob_{\fld} \dif \frac{\delta F}{\delta\fld} - \bs\cpl\act \\
\bs\cpl \cdot \dif \frac{\delta F}{\delta\fld} - \mob_n \act
\end{pmatrix} \\
& = \left[ \mob_{\fld} \dif \frac{\delta F}{\delta\fld} - \act\, \bs\cpl \right] \cdot \frac{\pd }{\pd a} \dif \frac{\delta F}{\delta\fld},
\end{align*}
and once again the contribution from the fuel dynamics vanishes.
Obviously, the same happens in the second term in the RHS of Eq.~\eqref{eq:ForcesPartialDerivative}.
The complete generalized force is
\begin{equation*}
G(t) = - \frac{1}{2T} \int_V \left\{ \left[ \bs\crt + \mob_{\fld} \dif \frac{\delta F}{\delta \fld} - \act\, \bs\cpl \right]
\cdot \dif \frac{\pd}{\pd a} \frac{\delta F}{\delta\fld} + \mathcal O(T) \right\} \dpos \,\D t',    
\end{equation*}
the same result we would have obtained if we used the dynamic action of only the system of interest from the beginning.
We interpret this result as a consequence of the fact that the active system dynamics can be solved without solving the fuel dynamics.
The term containing $\bs J$ can be integrated by parts using the divergence theorem, and we arrive at
\begin{equation} \label{eq:GeneralizedForce}
G(t) = - \frac{1}{2T} \int_V \Bigg\{ \left[ \mob_{\fld} \dif \frac{\delta F}{\delta \fld} - \act\, \bs\cpl \right] \cdot \dif \pd_a \frac{\delta F}{\delta\fld}
+ \dot \fld \pd_a \frac{\delta F}{\delta\fld} + \mathcal O(T) \Bigg\} \dpos,    
\end{equation}
and its functional derivative
\begin{equation} \label{eq:GeneralizedForceDerivated}
\frac{\delta G(t)}{\delta a(t')} = - \frac{1}{2T} \delta(t-t') \int_V \Bigg\{ \mob_{\fld} \left[ \dif \pd_a \frac{\delta F}{\delta\fld} \right]^2
+ \left[ \mob_{\fld} \dif \frac{\delta F}{\delta \fld} - \act\, \bs\cpl \right] \cdot \dif \pd_a^2 \frac{\delta F}{\delta\fld}
+ \dot\fld \pd_a^2 \frac{\delta F}{\delta\fld} + \mathcal O(T) \Bigg\} \dpos.  
\end{equation}

\section{Small noise expansion} \label{sec:smallNoiseExpansion}

In this section, we introduce the basics of the small noise expansion, a field approximation for low noise strength that allows for analytic results.
We start from the dynamics of $\fld$ (Eqs.~(3) and~(4) of the main text), which reads
\begin{subequations} \label{eq:phiDynamics}
\begin{align} \label{eq:continuityEquation}
\dot\fld + \dif \cdot & \bs\crt = 0 \\
\label{eq:currentDynamics}
& \bs\crt = - \mob_\fld \dif \frac{\delta F}{\delta\fld} + \bs\cpl \act + T^{1/2} \bs\noi_\fld + T \bs\spu_\fld,
\end{align}
\end{subequations}
with noise correlations given by (Eq.~(5) of the main text)
\begin{align} \label{eq:noiseCorrelations_pos&t}
\Avg{ \noi_\fld^\alpha(\pos,t) \noi_\fld^{\alpha'}(\pos',t') } = 2 \mob_\fld \delta^{\alpha,\alpha'} \delta(\pos-\pos') \delta(t-t')
\end{align}
and $\alpha$ and $\alpha'$ represent each of the $d$ dimensions.

Note that the global density of the system,
\begin{equation} \label{eq:GlobalDensity}
\bar\fld \dfn \frac{1}{V} \int_V \fld(\pos,t) \dpos,
\end{equation}
is a conserved quantity in the dynamics, which is easily proven by taking the time derivative of this definition and using Eq.~\eqref{eq:continuityEquation}.
Thus, as long as we stay in the homogeneous state of the system, and as long as the noises are small enough to only have perturbative effects around this homogeneous state, we can carry out a small noise expansion in $\fld$, as in
\begin{equation} \label{eq:SmallNoiseExpansion}
\fld(\pos,t) = \fld_0(\pos,t) + T^{1/2} \fld_1(\pos,t) + T \fld_2(\pos,t) + \mathcal O(T^{3/2}).
\end{equation}
The reason we use $T^{1/2}$ as the perturbative parameter here is that it is precisely the pre-factor of the noise $\bb\Lambda$ in Eq.~\eqref{eq:currentDynamics}, thus serving as a proxy to the noise strength.

Since we want Eq.~\eqref{eq:SmallNoiseExpansion} to represent small effects around the homogeneous state, for consistency we must have
\begin{subequations} \label{eq:smallNoiseConsistency}
\begin{align}
\fld_0(\pos,t) & = \bar\fld \\
\shortintertext{and}
\int_V \fld_1(\pos,t) \dpos & = 0 = \int_V \fld_2(\pos,t) \dpos,
\end{align}
\end{subequations}
and the same for higher orders.
Then, the space integral of $\fld^n$, where $n$ is an integer, becomes
\begin{align}
\notag
\int_V \fld^n \dpos
& = \int_V \left[ \fld_0^n + n \fld_0^{n-1} \left( T^{1/2} \fld_1 + T \fld_2 \right) + \frac{n(n-1)}{2} T \fld_0^{n-2} \fld_1^2 \right] \dpos + \mathcal O(T^{3/2}) \\
\label{eq:SNEpowerIntegral}
& = V \bar\fld^n + \frac{n(n-1)}{2} T \bar\fld^{n-2} \int_V \fld_1^2 \dpos + \mathcal O(T^{3/2}),
\end{align}
where we used Eq.~\eqref{eq:smallNoiseConsistency} in the second line.
Thus, any integer power of $\fld$ is approximated by $\fld_1^2$ to linear order in $T$.
Similarly, the space integral of $(\dif\fld)^n$ for $n \geq 2$ becomes
\begin{align} \label{eq:SNEgradientIntegral}
\int_V (\dif\fld)^n \dpos = T^{n/2} \int_V (\dif \fld_1)^n \dpos + \mathcal O(T^{(n+1)/2}).
\end{align}
Thus, only a term proportional to $(\dif\fld)^2$ contributes linearly in $T$.

To proceed, define the Fourier transform
\begin{equation} \label{eq:FourierTransform}
\tilde\fld_1(\mom,t) \dfn \frac{1}{\sqrt V} \int_V \fld_1(\pos,t) e^{-i \mom \cdot \pos} \dpos
\quad \Rightarrow \quad
\fld_1(\pos,t) = \frac{1}{\sqrt V} \sum_{\mom} \tilde\fld_1(\mom,t) e^{i \mom \cdot \pos}.
\end{equation}
Since $\fld_1$ is real, we have $\tilde\fld_1(-\mom,t) = \tilde\fld_1^*(\mom,t)$.
Consequently,
\begin{align} \label{eq:FourierQuadratics}
\int_{V} \fld_1^2(\pos,t) \dpos = \sum_{\mom} |\tilde\fld_1(\mom,t)|^2 \quad \text{and} \quad \int_V \big( \dif\fld_1(\pos,t) \big)^2 \dpos = \sum_{\mom} \mom^2 |\tilde\fld_1(\mom,t)|^2.
\end{align}
Now, consider an observable $O$ (possibly dependent on the external parameter $a$) written as
\begin{align}
O([\fld],a) = \int_V o(\phi,a) \dpos, \qquad o(\fld,a) = \sum_{n=2}^N c_n(a) \frac{\fld^n}{n} + \sum_{n=2}^{M} d_n(a) \frac{(\dif\fld)^n}{n},
\end{align}
where $c_n$ and $d_n$ are $\fld$-independent, while $N$ and $M$ are finite.
Every observable in this paper can be written in this form.
Equations~\eqref{eq:SNEpowerIntegral}, \eqref{eq:SNEgradientIntegral} and~\eqref{eq:FourierQuadratics} can be used to show that, up to first non-trivial order in the small noise expansion, we can write
\begin{align} \label{eq:SNEgenericObservable}
O([\fld],a) = O(\bar\phi,a) + \frac{T}{2} \sum_{\mom} \tilde o_1(a;\mom) |\tilde\fld_1(\mom,t)|^2 + \mathcal O(T^{3/2}),
\end{align}
where
\begin{align}
\tilde o_1(a;\mom) = \sum_{n=2}^N (n-1) \bar\fld^{n-2} c_n(a) + d_2(a) \mom^2.
\end{align}
Equation~\eqref{eq:SNEgenericObservable} allows us to calculate the average of any observable through the average $\avg{|\tilde\fld_1|^2}$.

\subsection{Steady-state averages} \label{sec:steadyStateAverages}

Our goal in this section is to calculate the steady-state average $\avg{\tilde\fld_1(\mom,t) \tilde\fld_1^*(\mom',t')}_\text{s}$, which serves to calculate every other steady-state average in the paper.
As such, in this section, we assume $a$ to be constant.

Much like the small noise expansion makes the observables quadratic in $\tilde\fld_1$ (see Eq.~\eqref{eq:SNEgenericObservable}), it also linearizes quantities such as $\delta F/\delta\fld$ and $\bs\cpl$ in $\tilde\fld_1$.
Therefore, to find the $\tilde\fld_1$ dynamics, we place the small noise expansion for $\fld$ (Eq.~\eqref{eq:SmallNoiseExpansion}) into the dynamics for $\fld$ (Eqs.~\eqref{eq:phiDynamics}), collect every term proportional to $T^{1/2}$ in both sides of the equation, and then take the Fourier transform defined in Eq.~\eqref{eq:FourierTransform}.
Doing so leads to
\begin{equation} \label{eq:phi1DynamicsMomentumSpace}
\dot{\tilde\fld}_1(\mom,t) = - \rlx(a;\mom) \tilde\fld_1(\mom,t) - i \mom \cdot \tilde{\bs\noi}_\fld(\mom,t),
\end{equation}
where the noise correlations in Fourier space reads
\begin{align} \label{eq:noiseCorrelations_mom&t}
\Avg{ \tilde\noi_\fld^\alpha(\mom,t) [\tilde\noi_\fld^{\alpha'}(\mom',t')]^* } = 2 \mob_\fld \delta^{\alpha,\alpha'} \delta_{\mom,\mom'} \delta(t-t'),
\end{align}
as derived from Eq.~\eqref{eq:noiseCorrelations_pos&t}.
In Eq.~\eqref{eq:phi1DynamicsMomentumSpace}, $\rlx(a;\mom)$ is the relaxation frequency of mode $\mom$ for a given model, only calculable once the free energy $F$ and the coupling $\bs C$ are specified.

Now, we define a new Fourier transform on the time variable,
\begin{equation} \label{eq:TimeFourierTransform}
\widetilde\fld_1(\mom,\omega) = \frac{1}{\sqrt{2\pi}} \int_{-\infty}^{\infty} \tilde\fld_1(\mom,t) e^{-i \omega t} \D t \quad \Rightarrow \quad \tilde\fld_1(\mom,t) = \frac{1}{\sqrt{2\pi}} \int_{-\infty}^{\infty} \widetilde\fld_1(\mom,\omega) e^{i \omega t} d\omega.
\end{equation}
It is important to note that such a Fourier transform on the time variable can only be done on $\fld_1$ (and not on the whole field $\fld$), because $\fld_1$ is expected to vanish as $t \to \pm \infty$.
Equation~\eqref{eq:phi1DynamicsMomentumSpace} can be shown to be equivalent to
\begin{equation} \label{eq:phi1DynamicsMomentumFrequencySpace}
\widetilde\fld_1(\mom,\omega) = - \frac{ i \mom \cdot \widetilde{\bs\noi}_\fld(\mom,\omega) }{ \rlx(a;\mom) + i \omega },
\end{equation}
whereas Eq.~\eqref{eq:noiseCorrelations_mom&t} is equivalent to
\begin{align} \label{eq:noiseCorrelations_mom&omega}
\Avg{ \widetilde{\noi}_\fld^{\alpha}(\mom,\omega) [\widetilde{\noi}_\fld^{\alpha'}(\mom',\omega')]^* } = 2\mob_\fld \delta^{\alpha,\alpha'} \delta_{\mom,\mom'} \delta(\omega-\omega').
\end{align}

Using the inverse transform of Eq.~\eqref{eq:TimeFourierTransform}, we get
\begin{align}
\notag
\Avg{ \tilde\fld_1(\mom,t) \tilde\fld_1^*(\mom',t') }_\text{s}(a)
& = \frac{1}{2\pi} \int_{-\infty}^{\infty} \int_{-\infty}^{\infty} \Avg{ \widetilde\fld_1(\mom,\omega)  \widetilde\fld_1^*(\mom',\omega') }_\text{s}(a) e^{i\omega t} e^{-i\omega' t'} d\omega d\omega' \\
\notag
& = \frac{1}{2\pi} \int_{-\infty}^{\infty} \int_{-\infty}^{\infty} \frac{ \Avg{ \left[ \mom \cdot \widetilde{\bs\noi}_\fld(\mom,\omega) \right] \left[ \mom' \cdot \widetilde{\bs\noi}_\fld(\mom',\omega') \right]^*  } }{ \left[ \rlx(a;\mom) + i \omega \right] \left[ \rlx(a;\mom') - i \omega' \right] } e^{i\omega t} e^{-i\omega' t'} d\omega d\omega' \\
\notag
& = \frac{1}{2\pi} 2\mob_\fld \mom^2 \delta_{\mom,\mom'} \int_{-\infty}^{\infty} \frac{ e^{i\omega [t-t']} }{ \rlx^2(a;\mom) + \omega^2 } d\omega \\
\label{eq:PhiPhiCorrelation}
& = \frac{\mob_\fld \mom^2}{ \rlx(a;\mom) } e^{-\rlx(a;\mom) |t-t'| } \delta_{\mom,\mom'}.
\end{align}
We used Eq.~\eqref{eq:phi1DynamicsMomentumFrequencySpace} in the second line and Eq.~\eqref{eq:noiseCorrelations_mom&omega} in the third line.

\subsection{Non-steady-state averages} \label{sec:nonSteadyStateAverages}

In this section, we assume $a$ to be time-dependent and intend to calculate the non-stationary average $\avg{\tilde\fld_1(\mom,t) \tilde\fld_1^*(\mom',t')}$.
We start from Eq.~\eqref{eq:phi1DynamicsMomentumSpace} with $a = a(t)$, i.e.,
\begin{equation} \label{eq:TDphi1Dynamics}
\dot{\tilde\fld}_1(\mom,t) = - \rlx(a(t);\mom) \tilde\fld_1(\mom,t) - i \mom \cdot \tilde{\bs\noi}_\fld(\mom,t),
\end{equation}
Now we have time-dependent dynamics, which means that a Fourier transform in the time variable does not help.
Instead, we define an auxiliary function for arbitrary $t_0$,
\begin{equation*}
\hat\fld_1(\mom,t) \dfn \tilde\fld_1(\mom,t) e^{\int_{t_0}^t \rlx(a(t');\mom) \D t'},
\end{equation*}
for which the differential equation, derived from Eq.~\eqref{eq:TDphi1Dynamics}, is
\begin{equation*}
\dot{\hat\fld}_1 = - i \mom \cdot \tilde{\bs\noi}_\fld e^{\int_{t_0}^t \rlx \D t'}.
\end{equation*}
This equation is easily integrated to give, after returning to $\tilde\fld_1$,
\begin{equation} \label{eq:TDphi1Integrated}
\tilde\fld_1(\mom,t) = e^{ - \int_{t_0}^t \rlx \D t_1 } \tilde\fld_1(\mom,t_0) - i \mom \cdot \int_{t_0}^{t} \tilde{\bs\noi}_\fld(\mom,t_2) e^{ - \int_{t_2}^{t} \rlx \D t_1 } \D t_2.
\end{equation}

From Eq.~\eqref{eq:TDphi1Integrated}, we can write
\begin{multline*}
\Avg{ \tilde\fld_1(\mom,t) \tilde\fld_1^*(\mom',t') } = e^{ - \int_{t_0}^t \rlx \D t_1 } e^{ - \int_{t_0}^{t'} \rlx' \D t_1' } \Avg{ \tilde\fld_1(\mom,t_0) \tilde\fld_1^*(\mom',t_0) } \\
+ i \mom' \cdot \int_{t_0}^{t'} \Avg{ \tilde\fld_1(\mom,t_0) \tilde{\bs\noi}_\fld^*(\mom',t_2') } e^{ - \int_{t_0}^t \rlx \D t_1 } e^{ - \int_{t_2'}^{t'} \rlx' \D t_1' } \D t_2' - i \mom \cdot \int_{t_0}^{t} \Avg{ \tilde\fld_1^*(\mom',t_0)  \tilde{\bs\noi}_\fld(\mom,t_2) } e^{ - \int_{t_0}^{t'} \rlx' \D t_1' } e^{ - \int_{t_2}^{t} \rlx \D t_1 } \D t_2 \\
+ \int_{t_0}^{t} \int_{t_0}^{t'} \Avg{ \left[ \mom \cdot \tilde{\bs\noi}_\fld(\mom,t_2) \right] \left[ \mom' \cdot \tilde{\bs\noi}_\fld^*(\mom',t_2') \right] } e^{ - \int_{t_2}^{t} \rlx \D t_1 } e^{ - \int_{t_2'}^{t'} \rlx' \D t_1' } \D t_2' \D t_2,
\end{multline*}
with $t \geq t'$.
The second line of this equation, representing the cross terms coming from the squaring of Eq.~\eqref{eq:TDphi1Integrated}, contain correlations between $\fld$ at $t_0$ and noises at later times (because $t_2 \geq t_0$).
Since the noise cannot influence the fields at previous times, these correlations are null.
The third line contain pure noise correlations, which are given in Eq.~\eqref{eq:noiseCorrelations_mom&t}, leading us to
\begin{align}
\label{eq:TDphi1phi1CorrelationGeneric}
\Avg{ \tilde\fld_1(\mom,t) \tilde\fld_1^*(\mom',t') }
& = e^{ - \int_{t_0}^t \rlx \D t_1 } e^{ - \int_{t_0}^{t'} \rlx' \D t_1' } \Avg{ \tilde\fld_1(\mom,t_0) \tilde\fld_1^*(\mom',t_0) }
+ 2\mob_\fld \mom^2 \delta_{\mom,\mom'} \int_{t_0}^{t'} e^{ - \int_{t_2}^{t} \rlx \D t_1 } e^{ - \int_{t_2}^{t'} \rlx' \D t_1' } \D t_2.
\end{align}

Now, to the specific result $\avg{ |\tilde\fld_1(\mom,t)|^2 }$, from which the non-steady-state average of every observable can be calculated in small noise (see Eq.~\eqref{eq:SNEgenericObservable}).
This can easily be obtained from Eq.~\eqref{eq:TDphi1phi1CorrelationGeneric} by setting $\mom' = \mom$ and $t'=t$, giving us
\begin{equation} \label{eq:TDphi1phi1Correlation}
\Avg{ |\tilde\fld_1(\mom,t)|^2 } = e^{ -2 \int_{t_0}^t \rlx(a(t_1);\mom) \D t_1 } \Avg{ |\tilde\fld_1(\mom,t_0)|^2 }
+ 2\mob_\fld \mom^2 \int_{t_0}^{t} e^{ -2 \int_{t_2}^t \rlx(a(t_1);\mom) \D t_1 } \D t_2.
\end{equation}
Thus, combining Eqs.~\eqref{eq:heats}, Eq.~\eqref{eq:SNEgenericObservable} and Eq.~\eqref{eq:TDphi1phi1Correlation}
we can calculate the heats $Q_\text{p}$ and $Q_\text{t}$ exactly in small noise.
Nevertheless, the integrals in Eq.~\eqref{eq:TDphi1phi1Correlation} cannot be solved analytically in most cases.
Thus, we solved them numerically for a linear protocol, a result used in Figs.~(3) and~(4) of the main text.

The one case where the integrals can be solved trivially is when $a$ is constant.
In this case, Eq.~\eqref{eq:TDphi1phi1Correlation} gives us
\begin{equation} \label{eq:TIphi1phi1Correlation}
\Avg{ |\tilde\fld_1(\mom,t)|^2 } = e^{ - 2\rlx(a;\mom) [t-t_0] } \Avg{ |\tilde\fld_1(\mom,t_0)|^2 } + \frac{\mob_\fld \mom^2}{\rlx(a;\mom)} \left[ 1 - e^{ - 2\rlx(a;\mom)[t-t_0] } \right], \qquad \text{if } \dot a = 0.
\end{equation}
This equation depicts a classic relaxation process.
No matter the average at $t_0$, this average always tends exponentially to a steady-state average when the system is left alone, i.e., when $\dot a(t) = 0$ for any $t$.
The relaxation time scale of each mode $\mom$ is given exactly by the inverse of the relaxation frequency $\rlx(a;\mom)$.
Additionally, if the system is already in its steady state at $t_0$, then the average in Eq.~\eqref{eq:TIphi1phi1Correlation} does not evolve, because the system is already relaxed.

\subsection{Integrated response functions} \label{sec:integratedResponseFunctions}

This section is dedicated to the calculation of the integrated response functions defined in Eqs.~(35) of the main text, which for a generic observable $O$ read
\begin{subequations} \label{eq:integratedResponseFunctions}
\begin{align}
\label{eq:zeta11}
\zeta_O^{(1,1)}(a) & = - \int_0^\infty t R_O^{(1)}(a;t) \D t, \\
\label{eq:zeta21}
\zeta_O^{(2,1)}(a) & = \frac{1}{2} \int_0^\infty t^2 R_O^{(1)}(a;t) \D t, \\
\shortintertext{and}
\label{eq:zeta22}
\zeta_O^{(2,2)}(a) &= \int_0^\infty \int_t^\infty t t' R_O^{(2)}(a;t,t') \D t' \D t,
\end{align} 
\end{subequations}
where $R_O^{(1)}$ and $R_O^{(2)}$ are defined in Eqs.~\eqref{eq:responseFunctions}.

Similarly to how the averages of observables simplify to the averages of $|\tilde\phi_1|^2$ in small noise, the same is true for $\zeta_O^{(m,n)}$.
More specifically, for a given observable given by Eq.~\eqref{eq:SNEgenericObservable}, its integrated response functions obey
\begin{align} \label{eq:SNEintegratedResponseFunctionGeneric}
\zeta_O^{(m,n)}(a) = \frac{T}{2} \sum_{\mom} \tilde o_1(a;\mom) \zeta_{|\tilde\fld_1|^2}^{(m,n)}(a;\mom) + \mathcal O(T^{3/2}).
\end{align}
Thus, we can focus on the integrated response functions of $|\tilde\fld_1|^2$.

Before we start, it must be noted that the response functions (e.g. Eq.~\eqref{eq:CorrelationFunction}) contain averages of products of observables at different times.
These can be written as averages of products of $|\tilde\fld_1|^2$ at different times, which can be simplified noting that $\tilde\fld_1$ has Gaussian statistics (see Eq.~\eqref{eq:phi1DynamicsMomentumFrequencySpace}) and using Isserli's theorem (also known as classical Wick's theorem).
Then, with the definitions of the two- and three-entry correlations of Eqs.~\eqref{eq:TwoEntryCorrelation}~and~\eqref{eq:ThreeEntryCorrelation}, we can write
\begin{equation} \label{eq:TwoEntryIsserlisTheorem}
\cor{ | \tilde\fld_1(\mom,t) |^2 | \tilde\fld_1(\mom_1,t_1) |^2 }_\text{s} = \left| \Avg{ \tilde\fld_1(\mom,t) \tilde\fld_1^*(\mom_1,t_1) }_\text{s} \right|^2 + \left| \Avg{ \tilde\fld_1(\mom,t) \tilde\fld_1^*(-\mom_1,t_1) }_\text{s} \right|^2
\end{equation}
and
\begin{multline} \label{eq:ThreeEntryIsserlisTheorem}
\cor{ | \tilde\fld_1(\mom,t) |^2 | \tilde\fld_1(\mom_1,t_1) |^2 | \tilde\fld_1(\mom_2,t_2) |^2 }_\text{s} = 2 \Re\bigg\{
\Avg{ \tilde\fld_1(\mom,t) \tilde\fld_1^*(\mom_1,t_1) }_\text{s} \Avg{ \tilde\fld_1(\mom_1,t_1) \tilde\fld_1^*(\mom_2,t_2) }_\text{s} \Avg{ \tilde\fld_1(\mom_2,t_2) \tilde\fld_1^*(\mom,t) }_\text{s} \\
+ \Avg{ \tilde\fld_1(-\mom,t) \tilde\fld_1^*(\mom_1,t_1) }_\text{s} \Avg{ \tilde\fld_1(\mom_1,t_1) \tilde\fld_1^*(\mom_2,t_2) }_\text{s} \Avg{ \tilde\fld_1(\mom_2,t_2) \tilde\fld_1^*(-\mom,t) }_\text{s} \\
+ \Avg{ \tilde\fld_1(\mom,t) \tilde\fld_1^*(-\mom_1,t_1) }_\text{s} \Avg{ \tilde\fld_1(-\mom_1,t_1) \tilde\fld_1^*(\mom_2,t_2) }_\text{s} \Avg{ \tilde\fld_1(\mom_2,t_2) \tilde\fld_1^*(\mom,t) }_\text{s} \\
+ \Avg{ \tilde\fld_1(\mom,t) \tilde\fld_1^*(\mom_1,t_1) }_\text{s} \Avg{ \tilde\fld_1(\mom_1,t_1) \tilde\fld_1^*(-\mom_2,t_2) }_\text{s} \Avg{ \tilde\fld_1(-\mom_2,t_2) \tilde\fld_1^*(\mom,t) }_\text{s}
\bigg\}.
\end{multline}

To start, we need the generalized force $G$ and its functional derivative $\delta G/\delta a$.
Writing the free energy in the form of Eq.~\eqref{eq:SNEgenericObservable},
\begin{align}
F([\fld],a) = F(\bar\fld;a) + \frac{T}{2} \sum_{\mom} \tilde f_1(a;\mom) | \tilde\fld_1(\mom,t) |^2 + \mathcal O(T^{3/2}),
\end{align}
the generalized force from Eq.~\eqref{eq:GeneralizedForce} can be written as
\begin{equation} \label{eq:GeneralizedForceMomentumSpace}
G(t_1) = - \frac{1}{2} \sum_{\mom_1} \pd_a \tilde f_1(a;\mom_1) \Bigg\{ \rlx(a,\mom_1) + \frac{1}{2} \frac{d}{dt_1} \Bigg\} | \tilde\fld_1(\mom_1,t_1) |^2 + \mathcal O(T^{1/2}),
\end{equation}
while its its functional derivative from Eq.~\eqref{eq:GeneralizedForceDerivated} becomes
\begin{equation} \label{eq:DifGeneralizedForceMomentumSpace}
\frac{\delta G(t_1)}{\delta a(t_2)} = - \frac{1}{2} \delta(t_1-t_2) \sum_{\mom_1} \left\{ \pd_a^2 \tilde f_1(a;\mom_1) \left[ \rlx(a,\mom_1) + \frac{1}{2} \frac{d}{dt_1} \right] + \pd_a \tilde f_1(a;\mom_1) \pd_a \rlx(a,\mom_1) \right\} | \tilde\fld_1(\mom_1,t_1) |^2 + \mathcal O(T^{1/2}).  
\end{equation}
As mentioned before, the $\mathcal O(T)$ terms in Eqs.~\eqref{eq:GeneralizedForce} and~\eqref{eq:GeneralizedForceDerivated} are dwarfed in our small noise treatment, and thus we put them together with the $\mathcal O(T^{1/2})$ terms in Eqs.~\eqref{eq:GeneralizedForceMomentumSpace} and~\eqref{eq:DifGeneralizedForceMomentumSpace}.
Then, the response functions follow straight from the results in Eqs.~\eqref{eq:CorrelationFunction} and~\eqref{eq:2ndResponseFunctionFromGeneralizedForce}.
For the first order with $t \geq t_1$, we get (omitting $\mathcal O(T^{1/2})$)
\begin{align}
\notag
R_{|\tilde\fld_1|^2}^{(1)}(a;\mom,t-t_1)
& = \cor{ |\tilde\fld_1(\mom,t)|^2 G(t_1) }_\text{s}(a) \\
\notag
& = - \frac{1}{2} \sum_{\mom_1} \pd_a \tilde f_1(a;\mom_1) \Bigg\{ \rlx(a,\mom_1) + \frac{1}{2} \frac{d}{dt_1} \Bigg\} \cor{ |\tilde\fld_1(\mom,t)|^2 | \tilde\fld_1(\mom_1,t_1) |^2 }_\text{s}(a) \\
\notag
& = - \sum_{\mom_1} \pd_a \tilde f_1(a;\mom_1) \Bigg\{ \rlx(a,\mom_1) + \frac{1}{2} \frac{d}{dt_1} \Bigg\} \left| \Avg{ \tilde\fld_1(\mom,t) \tilde\fld_1^*(\mom_1,t_1) }_\text{s}(a) \right|^2 \\
\notag
& = - \pd_a \tilde f_1(a;\mom) \Bigg\{ \rlx(a,\mom) + \frac{1}{2} \frac{d}{dt_1} \Bigg\} \left( \frac{\mob_\fld \mom^2}{ \rlx(a;\mom) } \right)^2 e^{-2\rlx(a;\mom) [t-t_1] } \\
\label{eq:phiphi1stResponseFunction}
& = - \pd_a \tilde f_1(a;\mom) \frac{2\mob_\fld^2 \mom^4}{ \rlx(a;\mom) } e^{-2\rlx(a;\mom) [t-t_1] },
\end{align}
where we used Eq.~\eqref{eq:GeneralizedForceMomentumSpace} in the second line, Eq.~\eqref{eq:TwoEntryIsserlisTheorem} in the third line, and Eq.~\eqref{eq:PhiPhiCorrelation} (the steady-state average $\avg{|\tilde\fld_1|^2}_\text{s}$ ) in the fourth line.
Similarly, the second order response function is (for $t \geq t_1 \geq t_2$)
\begin{multline} \label{eq:phiphi2ndResponseFunction}
R_{|\tilde\fld_1|^2}^{(2)}(a;\mom,t-t_1,t-t_2) = \big( \pd_a \tilde f_1(a;\mom) \big)^2 \frac{4\mob_\fld^3 \mom^6}{ \rlx(a;\mom) } e^{-2\rlx(a;\mom) [t-t_2] } \\
- \delta(t_1-t_2) \left[ 2\rlx(a,\mom) \pd_a^2 \tilde f_1(a;\mom) + \pd_a \tilde f_1(a;\mom) \pd_a \rlx(a,\mom) \right] \left( \frac{\mob_\fld \mom^2}{ \rlx(a;\mom) } \right)^2 e^{-2\rlx(a;\mom) [t-t_1] },
\end{multline}
where we also used Eqs.~\eqref{eq:ThreeEntryIsserlisTheorem} and~\eqref{eq:DifGeneralizedForceMomentumSpace}.

Next, we use the definitions in Eqs.~\eqref{eq:integratedResponseFunctions} to calculate the integrated response functions.
From Eq.~\eqref{eq:phiphi1stResponseFunction}, we get
\begin{align}
\notag
\zeta_{|\tilde\fld_1|^2}^{(1,1)}(a;\mom) 
& = - \int_0^\infty t R_{|\tilde\fld_1|^2}^{(1)}(a;\mom,t) \D t \\
\notag
& = \pd_a \tilde f_1(a;\mom) \frac{2\mob_\fld^2 \mom^4}{ \rlx(a;\mom) } \int_0^\infty t e^{-2\rlx(a;\mom) t } \D t \\
\label{eq:phi1phi1Zeta11}
& = \pd_a \tilde f_1(a;\mom) \frac{\mob_\fld^2 \mom^4}{ 2\rlx^3(a;\mom) },
\end{align}
and
\begin{align}
\notag
\zeta_{|\tilde\fld_1|^2}^{(2,1)}(a;\mom)
& = \frac{1}{2} \int_0^\infty t^2 R_{|\tilde\fld_1|^2}^{(1)}(a;\mom,t) \D t \\
\notag
& = - \pd_a \tilde f_1(a;\mom) \frac{\mob_\fld^2 \mom^4}{ \rlx(a;\mom) } \int_0^\infty t^2 e^{-2\rlx(a;\mom) t } \D t \\
\label{eq:phi1phi1Zeta21}
& = - \pd_a \tilde f_1(a;\mom) \frac{\mob_\fld^2 \mom^4}{ 4\rlx^4(a;\mom) }.
\end{align}
Finally, from Eq.~\eqref{eq:phiphi2ndResponseFunction}, we arrive at
\begin{align}
\notag
\zeta_{|\tilde\fld_1|^2}^{(2,2)}(a;\mom)
& = \int_0^\infty \int_t^\infty t t' R_{|\tilde\fld_1|^2}^{(2)}(a;t,t') \D t' \D t \\
\label{eq:phi1phi1Zeta22}
& = \left[ 3 \mob_\phi \mom^2 \big( \pd_a \tilde f_1(a;\mom) \big)^2  - 2\rlx(a,\mom) \pd_a^2 \tilde f_1(a;\mom) - \pd_a \tilde f_1(a;\mom) \pd_a \rlx(a,\mom) \right] \frac{\mob_\fld^2 \mom^4}{ 4\rlx^5(a;\mom) }.
\end{align}

\subsection{Active work in the relaxation process} \label{sec:posthumousHeat}

Our aim in this section is to prove Eq.~\eqref{eq:activePowerRelaxationIntegral}, which includes the active work in the relaxation process.
The active power $P$, when written in the form of Eq.~\eqref{eq:SNEgenericObservable}, is
\begin{equation} \label{eq:SNEactivePower}
P(t) = \frac{T}{2} \sum_{\mom} \tilde p_1(\mom) |\tilde\fld_1(\mom,t)|^2 + \mathcal O(T^{3/2}),
\end{equation}
where $\tilde p_1$ must be determined once $\bs\cpl$ is specified (see Eq.~\eqref{eq:CMBactivePowerMomentum} for the CMB case).
Thus, we can focus only on the relaxation of $|\tilde\fld_1|^2$, which is described in Eq.~\eqref{eq:TIphi1phi1Correlation} with $t_0 = \tau$ and $a = a_\tau$,
\begin{equation} \label{eq:phi1phi1Relaxation}
\Avg{ |\tilde\fld_1(\mom,t)|^2 } = \Avg{ |\tilde\fld_1(\mom,\tau)|^2 }_\text{s} + \left( \Avg{ |\tilde\fld_1(\mom,\tau)|^2 } - \Avg{ |\tilde\fld_1(\mom,\tau)|^2 }_\text{s} \right) e^{ - 2\rlx(a_\tau;\mom)[t-\tau] },
\end{equation}
where $\rlx$ is the relaxation frequency of the specific model, and we identified $\avg{ |\tilde\fld_1|^2 }_\text{s} = \mob_\fld \mom^2/\rlx$.
Note that the $t$-dependence in Eq.~\eqref{eq:phi1phi1Relaxation} is only found in the arguments of the exponentials.

Although DRT does not apply to the full relaxation process ($t \geq \tau$), it does apply to $t=\tau$, and thus we can use Eq.~(34) of the main text up to first order to write
\begin{equation} \label{eq:DRTphi1phi1}
\Avg{ |\tilde\fld_1(\mom,\tau)|^2 } = \Avg{ |\tilde\fld_1(\mom,\tau)|^2 }_\text{s}(a_\tau) + \dot a(\tau) \zeta_{|\tilde\fld_1|^2}^{(1,1)}(a_\tau;\mom) + \mathcal O(\ddot a, \dot a^2),
\end{equation}
where $\zeta^{(1,1)}_{|\fld_1|^2}$ is the first-order integrated response function of $|\tilde\fld_1|^2$ (see Eq.~\eqref{eq:phi1phi1Zeta11}).

Combining Eqs.~(\ref{eq:SNEactivePower}-\ref{eq:DRTphi1phi1}), we get
\begin{align} \label{eq:ActivePowerRelaxationIntegral0}
\int_{\tau}^{\tau+\tau_\text{r}} \Avg{ P(t) } \D t
& = \frac{T}{2} \sum_{\mom} \tilde p_1(\mom) \left[ \Avg{ |\tilde\fld_1(\mom,\tau)|^2 }_\text{s} \tau_\text{r} + \dot a(\tau) \frac{ \zeta_{|\tilde\fld_1|^2}^{(1,1)}(a_\tau;\mom) }{2\rlx(a_\tau;\mom)} \left( 1 - e^{ - 2\rlx(a_\tau;\mom) \tau_\text{r} } \right) \right] + \mathcal O(T^{3/2}).
\end{align}
Now, looking at the total heat in Eq.~\eqref{eq:DRTtotalHeat0}, we see that we also need $\dot a(\tau)\zeta^{(2,1)}_P(a_\tau)$, which for the generic form of $P$ in Eq.~\eqref{eq:SNEactivePower} reads
\begin{align} \label{eq:SNEactivePowerZeta21}
\dot a(\tau)\zeta^{(2,1)}_P(a_\tau) = \dot a(\tau) \frac{T}{2} \sum_{\mom} \tilde p_1(\mom) \zeta_{|\tilde\fld_1|^2}^{(2,1)}(a_\tau;\mom) + \mathcal O(T^{3/2}).
\end{align}
Thus, summing Eqs.~\eqref{eq:ActivePowerRelaxationIntegral0} and~\eqref{eq:SNEactivePowerZeta21}, we find
\begin{multline} \label{eq:activePowerRelaxationIntegral1}
\dot a(\tau)\zeta^{(2,1)}_P(a_\tau) + \int_{\tau}^{\tau+\tau_\text{r}} \Avg{ P(t) } \D t = \tau_\text{r} \frac{T}{2} \sum_{\mom} \tilde p_1(\mom) \Avg{ |\tilde\fld_1(\mom,\tau)|^2 }_\text{s} \\
+ \dot a(\tau) \frac{T}{2} \sum_{\mom} \tilde p_1(\mom) \left( \frac{ \zeta_{|\tilde\fld_1|^2}^{(1,1)}(a_\tau;\mom) }{2\rlx(a_\tau;\mom)} + \zeta_{|\tilde\fld_1|^2}^{(2,1)}(a_\tau;\mom) \right) \\
- \dot a(\tau) \frac{T}{2} \sum_{\mom} \tilde p_1(\mom) \frac{ \zeta_{|\tilde\fld_1|^2}^{(1,1)}(a_\tau;\mom) }{2\rlx(a_\tau;\mom)} e^{ - 2\rlx(a_\tau;\mom) \tau_\text{r} } + \mathcal O(T^{3/2}).
\end{multline}

Comparing Eqs.~\eqref{eq:phi1phi1Zeta11} and~\eqref{eq:phi1phi1Zeta21}, we can see that $\zeta_{|\tilde\fld_1|^2}^{(1,1)} = - 2 \rlx \zeta_{|\tilde\fld_1|^2}^{(2,1)}$.
Therefore, the second term on the RHS of Eq.~\eqref{eq:activePowerRelaxationIntegral1} is null.
This leaves us with two terms: the first scales linearly with $\tau_\text{r}$, while the last decreases exponentially.
In fact, since we assumed $\tau_\text{r}$ to be the time required for the system to reach its steady state, then we must have $\rlx(a_\tau;\mom) \tau_\text{r} \gg 1$ for every $\mom$ by definition.
Thus, the last term can be neglected, leading us to
\begin{align} \label{eq:activePowerRelaxationIntegral2}
\dot a(\tau)\zeta^{(2,1)}_P(a_\tau) + \int_{\tau}^{\tau+\tau_\text{r}} \Avg{ P(t) } \D t = \tau_\text{r} \Avg{ P }_\text{s}(a_\tau),
\end{align}
where we identified $\Avg{ P }_\text{s}$ in the first term on the RHS of Eq.~\eqref{eq:activePowerRelaxationIntegral1} (compare to Eq.~\eqref{eq:SNEactivePower}).
Equation~\eqref{eq:activePowerRelaxationIntegral2} is the same as Eq.~\eqref{eq:activePowerRelaxationIntegral}, proving the result.

In conclusion, the non-functional (but still process-dependent) part of the protocol heat (Eq.~\eqref{eq:protocolHeat}), namely $\dot a(\tau) B_\text{p}$, is compensated in the following relaxation process, and thus should not affect the optimization procedure.
It must be noted that, while proved here only for the first order in the small noise expansion, this result is true for all orders, as long as all quadratic correlation functions between the many orders of the expansion obey a relaxation process such as Eq.~\eqref{eq:phi1phi1Relaxation}, possibly with different relaxation frequencies.
Therefore, the expression for the total heat $Q_\text{t}$ assuming DRT (Eq.~\eqref{eq:DRTtotalHeat}) holds true for any strength of the noise.

\section{Chemical Model B} \label{sec:chemicalModelB}

This section is dedicated to the active field theory we employed in the main text to test our optimization procedures, which we call Chemical Model B (CMB).
It is defined by the dynamics of Eqs.~\eqref{eq:phiDynamics} with the following free energy and coupling:
\begin{align}
\label{eq:CMBfreeEnergy}
F & = \int_V \left( a \frac{\fld^2}{2} + b \frac{\fld^4}{4} + \kappa \frac{(\dif\fld)^2}{2} \right) \dpos \\
\shortintertext{and}
\label{eq:CMBcoupling}
\bs\cpl & = \gamma \dif\fld,
\end{align}
where $b$, $\kappa$ and $\gamma$ are constants.
For future reference, the active power (Eq.~(11) of the main text) is
\begin{align} \label{eq:CMBactivePower}
P = \frac{\act}{\mob_\fld} \int_V \left( \bs\cpl \cdot \bs\crt - \act \, \bs\cpl^2 \right) \dpos.
\end{align}

\subsection{Phase-separation \label{sec:CMBheterogeneous}}

In this section, we present the phase-separated solution to the one-dimensional CMB in its steady state.
At zero temperature, the dynamic equation for the field $\fld$ is, taken from Eq.~(17) of the main text,
\begin{equation} \label{eq:ZeroTemperatureFieldDynamics}
\dot\fld(x,t) = \mob_\fld \frac{\pd^2}{\pd x^2} \left[ \left( a - \frac{\gamma \act}{\mob_\fld} \right) \fld(x,t) + b\fld^3(x,t) - \kappa \frac{\pd^2 \fld(x,t)}{\pd x^2} \right].
\end{equation}

Since there are no thermal fluctuations, the steady state solution is easily found by setting $\dot\fld = 0$.
In that case, a possible solution of Eq.~\eqref{eq:ZeroTemperatureFieldDynamics}, when $a + b\bar\fld^2 < \gamma \act/\mob_\fld$, is
\begin{equation} \label{eq:HeterogeneousSolution}
\fld(x) = \fld_\text{het} \tanh\left( k_\text{het} x - A_\text{het} \right),
\end{equation}
where $\fld_\text{het} \dfn \sqrt{ (\gamma \act/\mob_\fld - a)/b }$, $k_\text{het} \dfn \sqrt{ (\gamma \act/\mob_\fld - a)/(2\kappa) }$ and
\[
A_\text{het} \dfn \frac{1}{2} \log\left( \frac{ \sinh\left[ \frac{k_\text{het} V}{2} \left( 1 - \frac{\bar\fld}{\fld_\text{het}} \right) \right] }{ \sinh\left[ \frac{k_\text{het} V}{2} \left( 1 + \frac{\bar\fld}{\fld_\text{het}} \right) \right] } \right).
\]
The field of Eq.~\eqref{eq:HeterogeneousSolution}, plotted in the inset of Fig.~2(b) of the main text, contains an interface between two regions of opposite densities $\phi_{\pm} = \pm \fld_\text{het}$, signaling a coexistence of phases.
While $\fld_\text{het}$ and $k_\text{het}$ are necessary for Eq.~\eqref{eq:HeterogeneousSolution} to be a solution of Eq.~\eqref{eq:ZeroTemperatureFieldDynamics}, $A_\text{het}$ only comes from normalization ( $\int_{-V/2}^{V/2} \fld(x) dx = V \bar\fld$ ), and it is only real when $a + b\bar\fld^2 < \gamma \act/\mob_\fld$.
The curve $\tilde a = a - \gamma \act/\mob_\fld = - b \bar\fld^2$ is what defines the first order transition line in Fig.~2(a) of the main text.

The local active power $p$ (see Eq.~(20) of the main text) can easily be obtained for this heterogeneous solution by using Eq.~\eqref{eq:HeterogeneousSolution},
\begin{align}
\notag
p(x) 
& = \frac{\act}{\mob_\phi} \left[ \bs\cpl \cdot \bs\crt - \act\, \bs\cpl^2 \right] \\
\notag
& = \frac{\gamma \act}{\mob_\fld} \left[ \frac{1}{2} \frac{\pd \fld^2(x)}{\pd t} - \gamma \act \left( \frac{\pd\fld(x)}{\pd x} \right)^2 \right] \\
\label{eq:HeterogeneousLocalActivePower}
& = - \frac{(k_\text{het} \fld_\text{het} \gamma \act)^2}{\mob_\fld} \text{sech}^4\left( k_\text{het} x - A_\text{het} \right).
\end{align}
The first term inside the square brackets, coming from $\bs\cpl \cdot \bs\crt$, vanishes, owning to the passive-like nature of CMB.
Equation~\eqref{eq:HeterogeneousLocalActivePower} is also plotted on the inset of Fig.~2(b) of the main text.

From Eq.~\eqref{eq:HeterogeneousLocalActivePower}, it is straightforward to obtain the global active power.
In the homogeneous state $\fld(x) = \bar\fld$, the only valid steady-state solution at zero temperature when $a + b\bar\fld^2 > \gamma \act/\mob_\fld$, the local and global active power are evidently zero.
Thus, we can write the global active power in steady state as a function of $\act$ as in
\begin{equation} \label{eq:HeterogeneousGlobalActivePower}
P(\act) = \begin{cases}
0, & a + b\bar\fld^2 > \gamma \act/\mob_\fld; \\
\displaystyle{ - \frac{(\fld_\text{het} \gamma \act)^2}{\mob_\fld} \frac{2k_\text{het}}{3} \frac{ \cosh(2 k_\text{het} V) + 6 \cosh(k_\text{het} V) \cosh(2A_\text{het}) + 5 }{ [ \cosh(k_\text{het} V) + \cosh(2A_\text{het}) ]^3 } \sinh(k_\text{het} V) },
& a + b\bar\fld^2 < \gamma \act/\mob_\fld.
\end{cases}
\end{equation}
Equation~\eqref{eq:HeterogeneousGlobalActivePower} is plotted on Fig.~2(b) of the main text.

\subsection{Steady-state averages in the homogeneous phase} \label{sec:CMBsteadyStateAverages}

In this section we obtain the relevant steady-state averages of CMB, using the results of Sec.~\ref{sec:smallNoiseExpansion}.
To start, we note that the free energy of Eq.~\eqref{eq:CMBfreeEnergy}, when expressed in the form of Eq.~\eqref{eq:SNEgenericObservable}, becomes
\begin{align} \label{eq:CMBfreeEnergyMomentum}
F = \left( a \frac{\bar\fld^2}{2} + b \frac{\bar\fld^4}{4} \right) V + \frac{T}{2} \sum_{\mom} \tilde f_1(a;\mom) |\tilde\fld_1(\mom,t)|^2 + \mathcal O(T^{3/2}),
\end{align}
where
\begin{align} \label{eq:CMBstructureFactor}
\tilde f_1(a;\mom) = a + 3b \bar\fld^2 + \kappa \mom^2.
\end{align}
Furthermore, the coupling of Eq.~\eqref{eq:CMBcoupling}, expanded in small noise with Eq.~\eqref{eq:SmallNoiseExpansion} and Fourier-transformed with Eq.~\eqref{eq:FourierTransform}, becomes
\begin{align} \label{eq:CMBcouplingMomentum}
\tilde{\bs\cpl} = T^{1/2} i \mom \gamma \tilde\fld_1 + \mathcal{O}(T).
\end{align}
Consequently, the active power (Eq.~\eqref{eq:CMBactivePower}) also takes the form of Eq.~\eqref{eq:SNEgenericObservable},
\begin{align} \label{eq:CMBactivePowerMomentum}
P = \frac{T}{2} \frac{\gamma \act}{\mob_\fld} \sum_{\mom} \left( \frac{d}{dt} - 2 \gamma \act \mom^2 \right) |\tilde\fld_1(\mom,t)|^2 + \mathcal O(T^{3/2}).
\end{align}

Using Eqs.~\eqref{eq:CMBfreeEnergyMomentum} and~\eqref{eq:CMBcouplingMomentum} to find the time evolution of $\tilde\fld_1$ leads to linear dynamics of Eq.~\eqref{eq:phi1DynamicsMomentumSpace}, with relaxation frequency given by
\begin{align} \label{eq:CMBrelaxationFrequency}
\rlx(a;\mom) = \mob_\fld \mom^2 \left( \tilde f_1(a;\mom) - \gamma \act/\mob_\fld \right).
\end{align}
The smallest relaxation frequency of the system is
\begin{align}
\Omega \dfn \rlx(a_\text{min};k_\text{min}),
\end{align}
where $a_\text{min}$ is the smallest value $a$ reaches in the manipulation process and $k_\text{min} = 2\pi/V^{1/d}$.
Note that, as the volume $V$ increases, the lowest wave number $k_\text{min}$ goes to zero, such that the relaxation time scale $\Omega^{-1}$ diverges.
This is a consequence of the conservation of the field $\fld$, leading to the existence of so-called hydrodynamic modes.
To explore this further, we studied a non-conserved active field theory following model-A dynamics in Sec.~\ref{sec:ScalingModelA}, where we show that there are no divergence of natural time scales in the thermodynamic limit.

Finally, the relevant steady-state averages can easily be obtained by using the result for $\avg{|\tilde\fld_1|^2}_\text{s}$.
More specifically, by taking the average of $F$, $\pd_a F$ and $P$ in Eqs.~\eqref{eq:CMBfreeEnergyMomentum} and~\eqref{eq:CMBactivePowerMomentum} and employing Eq.~\eqref{eq:PhiPhiCorrelation}, we get
\begin{subequations} \label{eq:CMBsteadyStateAverages}
\begin{align}
\Avg{F}_\text{s}(a) & = \left( a \frac{\bar\fld^2}{2} + b \frac{\bar\fld^4}{4} \right) V + \frac{T}{2} \sum_{\mom} \tilde f_1(a;\mom) \frac{\mob_\fld \mom^2}{\rlx(a;\mom)} + \mathcal O(T^{3/2}), \\
\Avg{\pd_a F}_\text{s}(a) & = \frac{\bar\fld^2}{2} V + \frac{T}{2} \sum_{\mom} \frac{\mob_\fld \mom^2}{\rlx(a;\mom)} + \mathcal O(T^{3/2}), \\
\shortintertext{where we used $\pd_a \tilde f_1 = 1$ from Eq.~\eqref{eq:CMBstructureFactor}, and}
\Avg{P}_\text{s}(a) & = - T \frac{\gamma^2 \act^2}{\mob_\fld} \sum_{\mom} \mom^2 \frac{\mob_\fld \mom^2}{\rlx(a;\mom)} + \mathcal O(T^{3/2}).
\end{align}
\end{subequations}

\subsection{Integrated response functions in the homogeneous phase} \label{sec:CMBintegratedResponseFunctions}

This section is dedicated to the calculation of the integrated response functions within CMB.
Having already expressed $F$ and $P$ in the form of Eq.~\eqref{eq:SNEgenericObservable}, we can directly use Eq.~\eqref{eq:SNEintegratedResponseFunctionGeneric} (which relates $\zeta_O^{(m,n)}$ to $\zeta^{(m,n)}_{|\tilde\fld_1|^2}$) to calculate the relevant integrated response functions.

Looking at protocol and total heats (Eqs.~\eqref{eq:DRTheats} and~\eqref{eq:heatFactors}), we see that we need $\zeta_F^{(1,1)}$, $\zeta_{\pd_aF}^{(1,1)}$, $\zeta_P^{(1,1)}$, $\zeta_P^{(2,1)}$ and $\zeta_P^{(2,2)}$.
Thus, from Eq.~\eqref{eq:SNEintegratedResponseFunctionGeneric}, the free energy (Eq.~\eqref{eq:CMBfreeEnergyMomentum}), and $\zeta^{(1,1)}_{|\tilde\fld_1|^2}$ (Eq.\eqref{eq:phi1phi1Zeta11}), we get
\begin{subequations} \label{eq:CMBintegratedResponseFunctions}
\begin{align} 
\notag
\zeta_F^{(1,1)}(a) & = \frac{T}{2} \sum_{\mom} \tilde f_1(a;\mom) \zeta^{(1,1)}_{|\tilde\fld_1|^2}(a;\mom) + \mathcal O(T^{3/2}) \\
\label{eq:CMBzetaF11}
& = \frac{T}{2} \sum_{\mom} \tilde f_1(a;\mom) \frac{\mob_\fld^2 \mom^4}{ 2\rlx^3(a;\mom) } + \mathcal O(T^{3/2}) \\
\shortintertext{and}
\label{eq:CMBzetaDF11}
\zeta_{\pd_a F}^{(1,1)}(a) & = \frac{T}{2} \sum_{\mom} \frac{\mob_\fld^2 \mom^4}{ 2\rlx^3(a;\mom) } + \mathcal O(T^{3/2}),
\end{align}
where we used $\pd_a \tilde f_1 = 1$ (see Eq.~\eqref{eq:CMBstructureFactor}), while $\rlx$ is given in Eq.~\eqref{eq:CMBrelaxationFrequency}.
Similarly, we can use Eq.~\eqref{eq:SNEintegratedResponseFunctionGeneric}, the active power (Eq.~\eqref{eq:CMBactivePowerMomentum}),  and $\zeta^{(m,n)}_{|\tilde\fld_1|^2}$ (Eqs.~(\ref{eq:phi1phi1Zeta11}--\ref{eq:phi1phi1Zeta22})) to obtain
\begin{align}
\notag
\zeta_P^{(1,1)}(a) & = \frac{T}{2} \frac{\gamma \act}{\mob_\fld} \sum_{\mom} \left( -2\rlx(a;\mom) - 2\gamma \act \mom^2 \right) \zeta^{(1,1)}_{|\tilde\fld_1|^2}(a;\mom) + \mathcal O(T^{3/2}) \\
\label{eq:CMBzetaP11}
& = - T \gamma \act \sum_{\mom} \mom^2 \tilde f_1(a;\mom) \frac{\mob_\fld^2 \mom^4}{ 2\rlx^3(a;\mom) } + \mathcal O(T^{3/2}), \\
\label{eq:CMBzetaP21}
\zeta_P^{(2,1)}(a) & = T \gamma \act \sum_{\mom} \mom^2 \tilde f_1(a;\mom) \frac{\mob_\fld^2 \mom^4}{ 4\rlx^4(a;\mom) } + \mathcal O(T^{3/2}), \\
\shortintertext{and}
\label{eq:CMBzetaP22}
\zeta_P^{(2,2)}(a) & = - T \gamma \act \sum_{\mom} \mom^2 \tilde f_1(a;\mom) \frac{\mob_\fld^3 \mom^6}{ 2\rlx^5(a;\mom) } + \mathcal O(T^{3/2}).
\end{align}
\end{subequations}

\subsection{Scalings with volume and activity} \label{sec:CMBscalings}

In this section, we analyze the scalings of the quantities appearing in the heats (Eqs.~\eqref{eq:DRTheats}).
Although the Fourier transform defined in Eq.~\eqref{eq:FourierTransform} is discrete, it can be extrapolated in the limit of large system sizes.
More specifically, in the thermodynamic limit $V \to \infty$, the sums over Fourier modes can be cast as integrals.
For an arbitrary function $g(k)$, we can write
\begin{equation} \label{eq:SumToIntegral}
\sum_{\mom} g(k) \to \frac{V}{[2\pi]^d} \int_{k_\mathrm{min}}^{k_\mathrm{max}} g(k) \left( \frac{2\pi^{d/2}}{\Gamma(d/2)} k^{d-1} \right) dk,    
\end{equation}
where $d$ is the dimension of the system, $k_\mathrm{min} = 2\pi/V^{1/d}$ is the lower cutoff, $k_\mathrm{max} = \pi/\ell$ is the upper cutoff, $\ell$ is the lattice constant and $\Gamma$ is the standard gamma function.
This allows us to study the scaling of the steady-state averages and of the integrated response functions with $V$ when $V \to \infty$ and with $\act$ when $\act \to 0$.

For the rest of this section, we take $d=1$ to represent the 1D CMB of the main text.
Starting with the steady-state averages (Eqs.~\eqref{eq:CMBsteadyStateAverages}), after defining $\bar a \dfn a + 3b\bar\fld^2$, we get
\begin{subequations}
\begin{align}
\Avg{F}_\text{s}(a) & = \left[ a \frac{\bar\fld^2}{2} + b \frac{\bar\fld^4}{4} \right] V + \frac{T V}{2} \left[ \frac{k_\text{max}}{\pi} + \frac{1}{2} \frac{\gamma \act}{\{\mob_\fld \kappa [ \mob_\fld \bar a - \gamma \act  ]\}^{1/2}}\right] + \mathcal O(T^{3/2},V^0), \\
\Avg{\pd_a F}_\text{s}(a) & = \frac{\bar\fld^2}{2} V + \frac{T V}{2} \left[ \frac{1}{2} \frac{\mob_\fld}{\{\mob_\fld \kappa [ \mob_\fld \bar a - \gamma \act  ]\}^{1/2}}\right] + \mathcal O(T^{3/2},V^0) \\
\shortintertext{and}
\Avg{ P }_\text{s}(a) & = - \frac{T \gamma^2 \act^2 V}{\mob_\fld \kappa} \left\{ \frac{k_\text{max}}{\pi} - \frac{1}{2} \left[ \frac{\mob_\fld \bar a - \gamma \act }{\mob_\fld \kappa} \right]^{1/2} \right\} + \mathcal O(T^{3/2},V^0).
\end{align}
\end{subequations}
From these, it is clear all these averages scale as $V$ in the thermodynamic limit.
We can also see that $\avg{ P }_\text{s} \sim \act^2$, giving it the exact same $V$ and $\act$ scalings as $P_0$ of Eqs.~\eqref{eq:heats}.

Next, we move on to the continuum limit of the integrated response functions (Eqs.~\eqref{eq:CMBintegratedResponseFunctions}), which leads to
\begin{subequations} \label{eq:CMBintegratedResponseFunctionsContinuous}
\begin{align}
\zeta_F^{(1,1)}(a) & = \frac{T \mob_\fld V}{ 4 [ \mob_\fld \bar a - \gamma\act ]^3 } \left\{ \frac{\mob_\fld \bar a}{\pi k_\text{min}} - \frac{3}{16} \Big[ 4 \mob_\fld \bar a + \gamma\act \Big] \left[ \frac{\mob_\fld \kappa}{\mob_\fld \bar a - \gamma\act} \right]^{1/2} \right\} + \mathcal O(T^{3/2},V^0), \\
\zeta_{\pd_a F}^{(1,1)}(a) & = \frac{T \mob_\fld^2 V}{ 4 [ \mob_\fld \bar a - \gamma\act ]^3 } \left\{ \frac{1}{\pi k_\text{min}} - \frac{15}{16} \left[ \frac{\mob_\fld \kappa}{\mob_\fld \bar a - \gamma\act} \right]^{1/2} \right\} + \mathcal O(T^{3/2},V^0), \\
\zeta_{P}^{(1,1)}(a) & = - \frac{T \gamma\act V}{32} \frac{\mob_\fld \Big[4\mob_\fld\bar a - \gamma\act \Big]}{\sqrt{\mob_\fld\kappa [\mob_\fld \bar a - \gamma\act]^5 } } + \mathcal O(T^{3/2},V^0), \\
\zeta_{P}^{(2,1)}(a) & = \frac{T \gamma\act \mob_\fld V}{4 [ \mob_\fld \bar a - \gamma\act ]^4} \left\{ \frac{\mob_\fld \bar a}{\pi k_\text{min}} - \frac{15}{96} \Big[ 6 \mob_\fld \bar a + \gamma\act \Big] \left[ \frac{\mob_\fld \kappa}{\mob_\fld \bar a - \gamma\act} \right]^{1/2} \right\} + \mathcal O(T^{3/2},V^0) \\
\shortintertext{and}
\zeta_{P}^{(2,2)}(a) & = - \frac{T \gamma\act \mob_\fld^2 V}{2 [ \mob_\fld \bar a - \gamma\act ]^5} \left\{ \frac{\mob_\fld \bar a}{\pi k_\text{min}} - \frac{105}{768}\Big[ 8 \mob_\fld \bar a + \gamma\act \Big] \left[ \frac{\mob_\fld\kappa}{\mob_\fld \bar a - \gamma\act} \right]^{1/2} \right\} + \mathcal O(T^{3/2},V^0).
\end{align}
\end{subequations}
Because $k_\text{min} = 2\pi/V$, all of these (except for $\zeta_{P}^{(1,1)}$) scale as $V^2$ in the thermodynamic limit.
Additionally, $\zeta_F^{(1,1)}$ and $\zeta_{\pd_aF}^{(1,1)}$ scale as $\act^0$ in the low activity limit.
Consequently, $B_\text{p}$ and $m$ of Eqs.~\eqref{eq:heatFactors} also scale as $V^2$ and $\act^0$, which ultimately leads to the situations where the protocol heat $Q_\text{p}$ does not display a minimum, as demonstrated in Fig.~4 of the main text.

\subsection{Optimal protocols \label{sec:OptimalProtocols}}

The goal of this section is to find an expression for the optimal protocols of CMB.
The functional part of the heats in Eqs.~\eqref{eq:DRTheats} reads, after using the Lagargian in Eq.~\eqref{eq:DRTlagrangian},
\begin{align}
\int_0^\tau \left[ m(a(t)) \dot a^2(t) + \Avg{P}_\text{s}(a(t)) \right] \D t = \int_0^1 \left[ m(a(s)) \frac{\dot a^2(s)}{\tau} + \tau \Avg{P}_\text{s}(a(s)) \right] ds,
\end{align}
where we changed variables to $s = t/\tau$ and abused notation by writing $\dot a(s) = da/ds$.
Thus, in the dimensionless variable $s$, the Lagrangian can be written as
\begin{equation} \label{eq:Lagrangian}
L(a, \dot a) = m(a) \frac{\dot a^2}{\tau} + \tau \Avg{P}_\text{s}(a).
\end{equation}
This Lagrangian can be readily placed in the Euler-Lagrange equation,
\begin{equation} \label{eq:EulerLagrange}
\frac{\pd L}{\pd a_\text{op}} = \frac{d}{ds} \frac{\pd L}{\pd \dot a_\text{op}} \quad \Rightarrow \quad \ddot a_\text{op} = \frac{\tau^2 \pd_a \Avg{P}_\text{s}(a_\text{op}) - \dot a_\text{op}^2 \pd_a m(a_\text{op})}{2m(a_\text{op})},
\end{equation}
to furnish the optimal protocol for a given $\tau$.
Equation~\eqref{eq:EulerLagrange} is a non-linear second-order equation and thus computationally intensive.
Here, we show how to use the form of Eq.~\eqref{eq:Lagrangian} to more efficiently solve Eq.~\eqref{eq:EulerLagrange}.

Since the Lagrangian of Eq.~\eqref{eq:Lagrangian} does not depend explicitly on $s$, the optimal protocol exhibits a first integral of motion given by
\begin{equation} \label{eq:FirstIntegral}
I \dfn \frac{1}{\tau} \left( \dot a_\text{op} \frac{\pd L}{\pd \dot a_\text{op}} - L \right) = m(a_\text{op}) \frac{\dot a_\text{op}^2}{\tau^2} - \Avg{ P }_\text{s}(a_\text{op}),
\end{equation}
which can be rewritten as
\begin{equation} \label{eq:EulerLagrangeIntegrated}
 \left( \frac{1}{\tau} \frac{da_\text{op}}{ds} \right)^2 = \frac{I + \Avg{ P }_\text{s}(a_\text{op})}{m(a_\text{op})}.   
\end{equation}
Without further information, we might assume the optimal protocol to be monotonic, in which case Eq.~\eqref{eq:EulerLagrangeIntegrated} can be readily solved with a single integration to give $a_\text{op}(s)$ implicitly, as in
\begin{equation} \label{eq:OptimalProtocolImplicit_Monotonic}
\tau s = \left\vert \int_{a_0}^{a_\text{op}(s)} \sqrt{ \frac{m(a)}{I + \Avg{ P }_\text{s}(a)} } da \right\vert,
\end{equation}
once $I$ is determined as a function of $\tau$, $a_0$ and $a_\tau$ from
\begin{equation} \label{eq:OptimalDurationFromFirstIntegral_Monotonic}
\tau = \left\vert \int_{a_0}^{a_\tau} \sqrt{ \frac{m(a)}{I + \Avg{ P }_\text{s}(a)} } da \right\vert.
\end{equation}

It must be noted that, because the LHS of Eq.~\eqref{eq:EulerLagrangeIntegrated} is non-negative, there might be restrictions on the values of $I$ that limit the applicability of Eqs.~\eqref{eq:OptimalProtocolImplicit_Monotonic} and~\eqref{eq:OptimalDurationFromFirstIntegral_Monotonic}, arising from the properties of $\avg{P}_\text{s}$ and $m$.
For example, in the homogeneous phases of the systems treated in this paper, $m(a)$ is positive and monotonically decreasing, while $\avg{P}_\text{s}(a)$ is negative and monotonically increasing (see sections~\ref{sec:CMBsteadyStateAverages} and~\ref{sec:CMBintegratedResponseFunctions}).
With these assumptions, we can see that Eq.~\eqref{eq:EulerLagrangeIntegrated} implies a lower bound on the first integral, i.e., $I \geq - \min_a \Avg{ P }_\text{s}(a).$
This minimum happens at $a_\text{min} \dfn \min \{a_0, a_\tau \}$, and thus the lower bound on $I$ is $- \Avg{ P }_\text{s}(a_\text{min})$.
This, in turn, sets an upper bound on the protocol duration $\tau$ that can be put into Eq.~\eqref{eq:OptimalDurationFromFirstIntegral_Monotonic}, given by
\begin{equation} \label{eq:DurationUpperBound}
\hat\tau \dfn \left\vert\int_{a_0}^{a_\tau} \sqrt{ \frac{m(a)}{ \Avg{ P }_\text{s}(a) - \Avg{ P }_\text{s}(a_\text{min}) } } da \right\vert.
\end{equation}

Notwithstanding, $\tau$ is a predetermined parameter of the protocol and can not be limited in any way by the optimization procedure.
We can then infer that, when $\tau$ is bigger than $\hat\tau$, the assumption of monotonicity of the optimal protocol must be false, and its derivative must be zero at some point $s_1$.
Consequently, we need to integrate Eq.~\eqref{eq:EulerLagrangeIntegrated} with different signs in different regions of the integration range, invalidating Eqs.~\eqref{eq:OptimalProtocolImplicit_Monotonic} and~\eqref{eq:OptimalDurationFromFirstIntegral_Monotonic} in this scenario.
Given the assumptions on $\avg{P}_\text{s}$ and $m$, Eq.~\eqref{eq:EulerLagrange} reveals that $\ddot a_\text{op} > 0$, which means that $\dot a_\text{op}$ changes sign only once and we need to split the integration region in two.
Then, Eq.~\eqref{eq:EulerLagrangeIntegrated} shows that zero derivative can happen at $a_1 \dfn a_\text{op}(s_1)$ given implicitly by $I = - \avg{P}_\text{s}(a_1)$.
Moreover, since $I \geq - \avg{ P }_\text{s}(a_\text{min})$, we must have $a_1 \leq a_\text{min}$, otherwise $I$ would bigger than its lower bound.
Thus, the optimal protocol attains zero derivative only once when it equals $a_1$, which is below the original range $[a_0, a_\tau]$, and its initial derivative is always negative.

We can then conclude that, when $\tau > \hat\tau$, the optimal protocol is given implicitly by
\begin{equation} \label{eq:OptimalProtocolImplicit_Polytonic}
\tau s = \begin{cases}
- \displaystyle\int_{a_0}^{a_\text{op}(s)} \sqrt{ \dfrac{m(a)}{ \Avg{ P }_\text{s}(a) - \Avg{ P }_\text{s}(a_1) } } da, &
0 \leq s \leq s_1; \\
- \displaystyle\int_{a_0}^{a_1} \sqrt{ \dfrac{m(a)}{ \Avg{ P }_\text{s}(a) - \Avg{ P }_\text{s}(a_1) } } da
+ \displaystyle\int_{a_1}^{a_\text{op}(s)} \sqrt{ \dfrac{m(a)}{ \Avg{ P }_\text{s}(a) - \Avg{ P }_\text{s}(a_1) } } da, &
s_1 < s \leq 1;
\end{cases}
\end{equation}
after $a_1$ is determined as a function of $\tau$, $a_0$ and $a_\tau$ from
\begin{equation} \label{eq:OptimalDurationFromFirstIntegral_Polytonic}
\tau = - \int_{a_0}^{a_1} \sqrt{ \frac{m(a)}{ \Avg{ P }_\text{s}(a) - \Avg{ P }_\text{s}(a_1) } } da + \int_{a_1}^{a_\tau} \sqrt{ \frac{m(a)}{ \Avg{ P }_\text{s}(a) - \Avg{ P }_\text{s}(a_1) } } da.
\end{equation}
Some of the non-monotonic solutions in fact become smaller than $a_\text{1st}$, the first-order transition line, but merely because the small noise expansion around the homogeneous solution does not take into account the presence of the transition line.
To compensate, we forcefully alter a posteriori all optimal protocols to never cross the phase transition.

It is worth noting that we split the integration region in two because, in our case, $\ddot a_\text{op}(s)$ has a constant sign.
With different $a$-dependence on $\avg{P}_\text{s}$ and $m$, one might need to split the region in more parts.

\subsection{Iterative method of obtaining optimal durations for the optimal protocols} \label{sec:iterativeMethod}

In the main text, we estimated the position of the minimum $\tau_\text{t}$ of the total heat $Q_\text{t}$ for the optimal protocol by extrapolating the asymptotic results of the two master curves $\alpha_0$ and $\alpha_\infty$ into the intermediate region and finding their intersection.
There is, however, a more accurate numerical method of of finding $\tau_\text{t}$, which we describe in this section.

Our starting point is Eq.~(14) of the main text,
\begin{align} \label{eq:OptimalProcessDuration}
\tau_\text{t}^2 = \frac{\int_0^1 \dot a^2 m(a) ds}{P_0 + \int_0^1 \avg{P}_\text{s}(a) ds}.
\end{align}
This equation provides us with the optimal duration $\tau_\text{t}$ for a protocol that can be written as a function solely of $s = t/\tau$, thus it does not apply to the optimal protocols $a_\text{op}(\tau;s)$.
Nonetheless, we can pick an arbitrary duration $\tau_0$ and generate its associated protocol $a_\text{op}(\tau_0;s)$, as is done in Sec.~\eqref{sec:OptimalProtocols}.
This gives us a curve in the range $0\leq s\leq 1$, and despite this curve not being optimal for all values of $\tau$, we can still use it for any $\tau$.
Doing so gives us the total heat of this curve (Eq.~\eqref{eq:DRTtotalHeat}) and a well-defined minimum point given by the RHS of Eq.~\eqref{eq:OptimalProcessDuration}; call this $\tau_1$.
We can then repeat the steps, generating $a_\text{op}(\tau_1;s)$ and finding its minimum $\tau_2$.
This procedure converges to $\tau_\text{t}$ because the total heat always displays a global minimum.
Equation~(23) of the main text, which estimates $\tau_\text{t}$ from asymptotics, is in agreement with the procedure just described for all simulations done in this paper.

\subsection{Numerical simulations} \label{sec:numericalSimulations}

In this section, we present the basics of the numerical simulations shown in the main text, whose source code can be found in Ref.~\cite{simul}.
Working in 1D, we discretize time as $t = j \delta t$ where $j$ is a natural number, and space as $x = n \delta x$, where $n = 0, 1, \cdots, N_x-1$ and $N_x$ is the number of lattice points, obeying $ N_x \delta x = V$.
Labeling the field as $\fld_n^j$, we take its discrete Fourier transform $\mathcal F[\fld_n^j]_k \dfn \tilde\fld_k^j$ (where $k = 0, 2\pi/(N_x \delta x), \cdots \pi/\delta x$) and update it in each time step $\delta t$ using a pseudo-spectral scheme, such that space derivatives can be calculated exactly.
The current is calculated using the discretized form of Eq.~(17) of the main text,
\begin{equation} \label{eq:currentDiscretized}
\tilde\crt_k^j = - i \mob_\fld k \left\{ \left[ a(j\delta t) - \frac{\gamma \act}{\mob_\phi} + \kappa k^2 \right] \tilde\fld_k^j + b \mathcal F[(\fld_n^j)^3]_k \right\} + \sqrt{ \frac{2T\mob_\phi}{ \delta x \delta t } } \tilde\xi_k^j,
\end{equation}
where $\avg{\xi_k^j \xi_{k'}^{j'}} = \delta_{k,k'} \delta^{j,j'}$ and the spurious drift vanishes in 1D \cite{Markovich2021}.
Then, the field gets updated as
\begin{equation} \label{eq:fieldDiscretized}
\fld_n^{j+1} = \fld_n^j - \mathcal F^{-1}[ i k \tilde\crt_k^j ]_n \delta t,
\end{equation}
where we returned to position space because we need to calculate $\mathcal F[(\fld_n^j)^3]_k$ for Eq.~\eqref{eq:currentDiscretized} in every time step.
Note that Eqs.~\eqref{eq:currentDiscretized} and~\eqref{eq:fieldDiscretized} do not assume small noise, and thus are valid for any finite $T$.

Initially, we set $\fld_n^{j=0} = \bar\fld$ and let it evolve with $\dot a = 0$, until it reaches a steady state.
During the manipulation, the protocol and total heats are calculated as follows.
At each time step, we calculate the field at half time step,
\begin{equation} \label{eq:halfFieldDiscretized}
\fld_n^{j+1/2} = \fld_n^j - \mathcal F^{-1}[ i k \tilde \crt_k^j ]_n \delta t/2,
\end{equation}
and $\tilde\crt_k^{j+1/2}$ is calculated accordingly, with Eq.~\eqref{eq:currentDiscretized}.
The observables $\pd_a F$ and $P$ are retrieved from
\begin{align} \label{eq:pdFreeEnergyDiscretized}
\pd_a F^{j+1/2} & = \sum_n \frac{(\fld_n^{j+1/2})^2}{2} \delta x \\
\shortintertext{and} \label{eq:activePowerDiscretized}
P^{j+1/2}       & = \frac{\act}{\mob_\phi} \sum_k \left[i \gamma k \tilde\fld_k^{j+1/2}\right]^* \left[ \tilde\crt_k^{j+1/2} - i \gamma\act k \tilde\fld_k^{j+1/2}  \right].
\end{align}
Then, the external and active works are calculated from discretized Stratonovich integrals,
\begin{align} \label{eq:externalWorkDiscretized}
\mathcal W_\text{ext} = & \sum_{j=0}^{\tau/\delta t} \dot a\big((j+1/2)\delta t\big) \pd_a F^{j+1/2} \delta t \\
\shortintertext{and} \label{eq:activeWorkDiscretized}
\mathcal W_\text{act}^{j'} = & \sum_{j=0}^{j'} \left[ \mob_n V \act^2 + P^{j+1/2} \right] \delta t,
\end{align}
where the upper limit $j'$ in Eq.~\eqref{eq:activeWorkDiscretized} is chosen based on the heat one is trying to calculate.
The free energy, on the other hand, can be simply evaluated at any point with
\begin{equation} \label{eq:freeEnergyDiscretized}
F^j = \sum_n \left[ a(j\delta t) \frac{(\fld_n^j)^2}{2} + b \frac{(\fld_n^j)^4}{4} \right] \delta x + \kappa \sum_k k^2 \frac{\vert \tilde\fld_k^j \vert^2}{2}.
\end{equation}
Finally, the protocol and total heats, defined as fluctuating quantities in each noise realization, are obtained from
\begin{align} \label{eq:protocolHeatDiscretized}
\mathcal Q_\text{p} & = \mathcal W_\text{ext} + \mathcal W_\text{act}^{\tau/\delta t} - \left[ F^{\tau/\delta t} - F^0 \right] \\
\shortintertext{and} \label{eq:totalHeatDiscretized}
\mathcal Q_\text{t} & = \mathcal W_\text{ext} + \mathcal W_\text{act}^{[\tau + \tau_\text{r}]/\delta t} - \left[ F^{[\tau + \tau_\text{r}]/\delta t} - F^0 \right].
\end{align}
The heats shown in Figs.~3 and~5 of the main text are then mean values of Eqs.~\eqref{eq:protocolHeatDiscretized} and~\eqref{eq:totalHeatDiscretized} over many realizations of the same protocol, i.e., $Q_\text{p} = \avg{ \mathcal Q_\text{p} }$ and $Q_\text{t} = \avg{ \mathcal Q_\text{t} }$.

\begin{figure*}

\includegraphics[width=\linewidth]{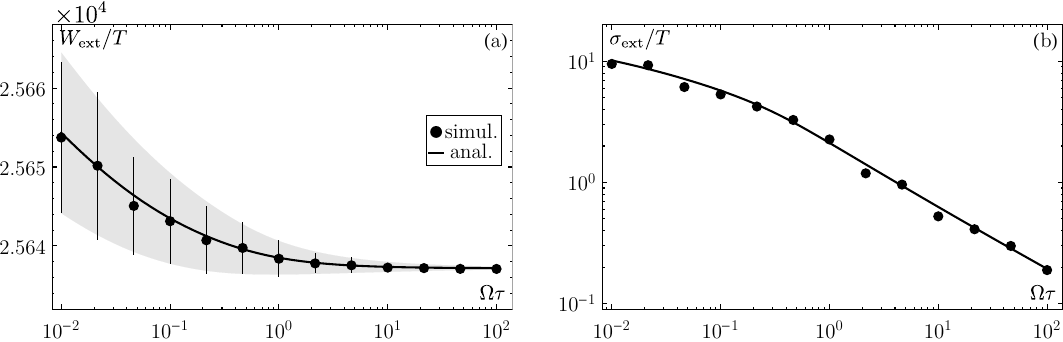}

\caption{\label{fig:externalWorkVariance}
Plots of the external work $W_\text{ext}$ and its variance $\sigma_\text{ext}$ as a function of protocol duration $\tau$.
Points represents the results from numerical simulations, while the solid line represents the analytical results in small noise.
Parameters: $a_0 = 1$, $a_\tau = 10$, $b = 1 = \kappa$, $\bar\phi = 0.5$, $\mob_\phi = \mob_n = \gamma = 1$, $V/\ell = 64$, $\act/T = 0.1$ and $F(\phi_0,a_0)/T = 50 V/\ell$.
(a) $W_\text{ext}$ vs $\tau$, where the horizontal lines represent the variance from numerical simulations, while the shaded region represent the variance calculated from Eq.~\eqref{eq:externalWorkVariance};
(b) $\sigma_\text{ext}$ vs $\tau$.
}

\end{figure*}

The energy flows fluctuate from one realization of the noise to the next, and we can compare the numerical fluctuations with analytical fluctuations from the exact result in small noise of Sec.~\ref{sec:nonSteadyStateAverages}.
For instance, in the case of the external work, we can write its variance as
\begin{equation} \label{eq:externalWorkVariance}
\sigma_\text{ext}^2 = 2 \int_0^\tau \int_0^t \dot a(t) \dot a(t') \sigma^2_{\pd_a F}(t,t') \D t' \D t,
\end{equation}
where
\begin{align}
\notag
\sigma^2_{\pd_a F}(t,t')
& \dfn \Avg{\pd_a F(t) \pd_a F(t')} - \Avg{ \pd_a F(t) } \Avg{ \pd_a F(t') } \\
\label{pdFreeEnergyVariance}
& = \frac{T^2}{2} \sum_{\mom,\mom'} \left| \Avg{ \tilde\fld_1(\mom,t) \tilde\fld_1^*(\mom',t') } \right|^2 + \mathcal O(T^{5/2}),
\end{align}
where the second line assumes small noise.
With help from Eq.~\eqref{eq:TDphi1phi1CorrelationGeneric} to express the non-steady-state average, we can evaluate the integral in Eq.~\eqref{eq:externalWorkVariance} numerically.
Figure~\ref{fig:externalWorkVariance}(a) shows that, while the variance from simulations are big compared to the variation of the mean between the edges of the plot, it still shows satisfactory agreement with the variance calculated from Eq.~\eqref{eq:externalWorkVariance}.
Figure~\ref{fig:externalWorkVariance}(b) corroborates this point further, showing the variance $\sigma_\text{ext}$ directly.
With considerable more effort and computing time, the same conclusions can be reached about the protocol and total heats.

\section{Scalings with volume and activity in model A \label{sec:ScalingModelA}}

\begin{figure*}

\includegraphics[width=\linewidth]{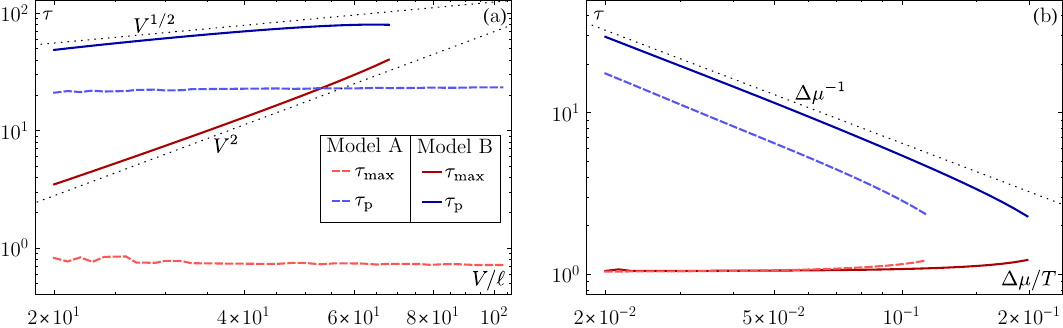}

\caption{\label{fig:ScalingModelA}
Plots comparing the volume and activity scalings of the maxima and minima of the protocol heat $Q_\text{p}$ in models B and A, obtained numerically from 
Parameters used: $a_0 = 1$, $a_\tau = 10$, $b = 10$, $\kappa = 1$, $\bar\fld = 0$, $\mob_\fld = \mob_n = \gamma = 1$ and $F(0.5,a_0)/T = 50 V/\ell$.
The dashed lines are merely guides for the annotated scalings.
(a) Durations $\tau_\text{max}$ and $\tau_\text{p}$ vs $V$, with $\act/T=0.02$;
(b) Durations $\tau_\text{max}$ and $\tau_\text{p}$ vs $\act/T$, with $V/\ell=10$.
}

\end{figure*}

In the main text, we offered a qualitative and quantitative analysis of the behavior of the maximum and minimum points of the process and total heats of CMB.
In this section, we study the scalings of a similar system without conservation law, following model A-like dynamics.
Its dynamic equation is
\begin{equation} \label{eq:CMAdynamics}
\dot\fld = - \mob_\fld \frac{\delta F}{\delta\fld} + C(\fld) \act + T^{1/2} \Lambda_\fld + T \spu_\fld, \qquad (\text{Model A})
\end{equation}
where all quantities are scalars, but defined in analogy to those of Model B in Eq.~\eqref{eq:currentDynamics}.
In specific, the coupling here should be chosen as $C(\phi) = \gamma \phi$.

Redoing all the calculations of Sec.~\ref{sec:CMBsteadyStateAverages} for the dynamics of Eq.~\eqref{eq:CMAdynamics} yields similar results.
One important difference is that, since the global density $\bar\fld$ is not constant in model A, the zeroth-order contribution to the small noise expansion ($\phi_0$ in Eq.~\eqref{eq:SmallNoiseExpansion}) may be time-dependent.
However, in the homogeneous phase, the minimum of the free energy is found at $\phi_0 = 0$, and all complications arising from a time-dependent $\phi_0$ disappear.

Thus, in the homogeneous phase, the main difference between models A and B is the natural frequency, which in model A reads
\begin{equation} \label{eq:CMARelaxationFrequency}
\rlx(a;\mom) = \mob_\fld [c(a;\mom) - \gamma \act/\mob_\fld], \qquad\qquad\qquad (\text{Model A})
\end{equation}
As is now obvious, considering the lowest Fourier mode $k_\text{min}=2\pi/V^{1/d}$, $1/\rlx(a;k_\text{min})$ in model A remains finite as $V \to \infty$, meaning that there is no divergence of relaxation time scales of a non-conserved field.
Consequently, the integrated response functions of model A (in analogy to Eqs.~\eqref{eq:CMBintegratedResponseFunctionsContinuous} of model B) scale linearly with $V$ in the thermodynamic limit.
The results of Secs.~\ref{sec:nonSteadyStateAverages} and~\ref{sec:posthumousHeat} also extend naturally to model A.

Figure~\ref{fig:ScalingModelA}(a) shows how the positions of maxima and minima for models B and A behave with $V$ (in 1D).
It is clear that, for model A, $\tau_\text{max}$ and $\tau_\text{p}$ tend to a constant as $V \to \infty$, meaning that changing the system's accessible volume has no impact in the shape of the protocol heat profiles.
This is not the case for model B, where the field conservation leads to a diverging natural time scale and, consequently, to the absence of maxima and minima for large enough $\tau$.
Figure~\ref{fig:ScalingModelA}(b) then shows the behavior with $\act$, in which there is not qualitative difference between the models because this is scaling is not related to field conservation.

\section{Evaluation of the spurious work \label{sec:StochasticIntegrals}}

In this section, we provide the correct way of evaluating the stochastic integral present in Eq.~(29) of the main text.
As mentioned before, this entails careful consideration of the Stratonovich discretization and the correlations between the noise and other factors.

Before approaching the integral itself, a discussion about the spurious drift is in order.
Within the matrix notation introduced in the main text, we can express the matrix $\bbnu$ as follows.
Let $\bb P$ be the square matrix where each line is an eigenvector of $\bb L$ and let $\bb D$ be the square diagonal matrix where each non-zero entry is the square root of an eigenvalue of $\bb L$.
Now, define $\bb M \dfn \bb P^{-1} \cdot \bb D \cdot \bb P$.
Consequently, $\bb M \cdot \bb M^\trp = \bb L$, meaning that $\bb M$ is essentially the ``square root'' of $\bb L$.
Then, defining the differential matrix operator
\begin{equation} \label{eq:DerivativeMatrix}
\frac{\delta}{\delta\bb n} \dfn \begin{pmatrix} \dif \frac{\delta}{\delta\fld} \\ \frac{\delta}{\delta n} \end{pmatrix},   
\end{equation}
the $u$-th element of the spurious drift matrix can be written as \cite{Markovich2021}
\begin{equation} \label{eq:SpuriousDrift}
\bbnu_u = \sum_{vw} \bb M_{uv} \frac{\delta}{\delta\bb n}_w \bb M_{wv},
\end{equation}
where the subscripts $u$, $v$ and $w$ run from $1$ to $d+1$.
The same matrix $\bb M$ can be used to redefine the noise $\bb\noi$ into an uncorrelated form.
More specifically, if we set
\begin{equation} \label{eq:UncorrelatedNoise}
\bb\noi(\pos,t) \dfn \bb M(\pos,t) \cdot \bbxi(\pos,t)
\end{equation}
and require that
\begin{equation} \label{eq:NoiseUncorrelations}
\Avg{ \bbxi(\pos,t) \bbxi^\trp(\pos',t') } = 2 \bb 1 \delta(\pos-\pos') \delta(t-t'),
\end{equation}
where $\bb 1$ is the identity matrix, then
\begin{equation} \label{eq:MatrixNoiseCorrelations}
\Avg{ \bb\noi(\pos,t) \bb\noi^\trp(\pos',t') } = 2 \bb L(\fld(\pos,t)) \delta(\pos-\pos') \delta(t-t')
\end{equation}
is guaranteed.
However, unlike $\bb\noi$, each element of the column matrix $\bbxi$ is uncorrelated from one another, i.e., $\Avg{\bbxi_u \bbxi_v} = 0$ for $u \neq v$.

Now, note that the time integral of $P_\text{spu}$ (see Eq.~(30) of the main text) between arbitrary times $t_1$ and $t_2$, can be written in more compact form using this matrix notation,
\begin{equation} \label{eq:qNoiseMatrixNotation}
 \Avg{ \int_{t_1}^{t_2} P_\text{spu}(t) \D t } = T^{1/2} \frac{\act}{\mob_\fld}  \Avg{\int_{t_1}^{t_2} \int_V \bb K^\trp \cdot \left[ \bb M \cdot \bbxi + T^{1/2} \bbnu \right] \dpos \D t },
\end{equation}
where we used Eq.~\eqref{eq:UncorrelatedNoise} and we defined
\begin{equation} \label{eq:MobilityMatrix}
\bb K \dfn \begin{pmatrix} - \bs\cpl \\ \mob_\fld \end{pmatrix}.
\end{equation}
Here we treat stochastic time integrals involving the uncorrelated noise $\bbxi$, such as the first term inside the integral in Eq.~\eqref{eq:qNoiseMatrixNotation}, for a generic matrix $\bb K$.
Only in the end we use the specific definition given by Eq.~\eqref{eq:MobilityMatrix}.

Now, we follow reference \cite{Cates2022}.
Consider the following time integral and its Stratonovich-discretized form:
\begin{equation} \label{eq:StratoIntegral}
 \Avg{\int_{t_1}^{t_2} \int_V g(\fld(\pos,t),n(\pos,t)) \bbxi_v(\pos,t) \dpos \D t } \dfn \Avg{ \sum_i \int_V g\Big(\fld^{i+1/2}(\pos),n^{i+1/2}(\pos)\Big) \bbxi_v^i(\pos) \dpos \,\delta t^{1/2} },
\end{equation}
where $g$ is a generic function of the fields $\fld$ and $n$, $\bbxi_v$ is the $v$-th element of the uncorrelated gaussian noise introduced in Eq.~\eqref{eq:UncorrelatedNoise}, $i$ indexes time steps (with size $\delta t$), and the fields on the RHS are evaluated at $i+1/2$ (the mid-point between $i$ and $i+1$), as dictated by the Stratonovich convention.
The noise correlations in discretized form read $\Avg{ \bbxi_u^i \bbxi_v^j } = 2 \delta_{uv} \delta^{ij}$ and, because of this, $\bbxi(\pos,t) \to \delta t^{-1/2} \bbxi^i(\pos)$, leading to the $\delta t^{1/2}$ factor in the RHS of Eq.~\eqref{eq:StratoIntegral}.

Because $\fld^{i+1/2}$ and $\bbxi_v^i$ in Eq.~\eqref{eq:StratoIntegral} are not evaluated at the same point in time, it is unclear how to calculate this average as is.
For small $\delta t$, we can Taylor expand $g$ as
\begin{equation*}
g(\fld^{i+1/2},n^{i+1/2}) \approx g(\fld^i,n^i) + \frac{\delta g}{\delta\bb n}^{i} \cdot \frac{1}{2} \left[ T^{1/2} \bb\noi^{i} \delta t^{1/2} \right],
\end{equation*}
where $\frac{\delta}{\delta\bb n}$ is the matrix defined in Eq.~\eqref{eq:DerivativeMatrix} and the term inside the square brackets came from the discretized form of the dynamics (see Eqs.~\eqref{eq:currentDiscretized} and~\eqref{eq:fieldDiscretized}), keeping only lowest order terms in $\delta t$.
Placing this in Eq.~\eqref{eq:StratoIntegral} gives
\begin{multline} \label{eq:StratoToIto}
 \Avg{\int_{t_1}^{t_2} \int_V g(\fld(\pos,t),n(\pos,t)) \bbxi_v(\pos,t) \dpos \D t } \approx \Avg{ \sum_i \int_V g(\fld^i,n^i) \bbxi_v^i(\pos) \dpos \,\delta t^{1/2} } \\
+ \Avg{ \sum_i \int_V \frac{\delta g}{\delta\bb n}^{i} \cdot \frac{1}{2} \left[ T^{1/2} \bb\noi^{i} \delta t^{1/2} \right] \bbxi_v^i(\pos) \dpos \,\delta t^{1/2} },
\end{multline}
The sums in the RHS of Eq.~\eqref{eq:StratoToIto} have all its terms evaluated at the same point in time --- these are known as Itô integrals.
The averages of the integrands are now calculable straightforwardly, and thus we proceed by commuting the averaging $\avg{\cdot}$ with the sum over time steps.
In the case of the first sum, $\fld^i$ and $\bbxi_u^i$ are statistically independent, and the fact that $\avg{\bbxi} = 0$ means that this whole sum vanishes.
After using $\bb\noi = \bb M \cdot \bbxi$, we arrive at
\begin{align*}
 \Avg{ \int_{t_1}^{t_2} \int_V g(\fld(\pos,t),n(\pos,t)) \bbxi_v(\pos,t) \dpos \D t }
& \approx \sum_i \Avg{ \int_V \frac{\delta g}{\delta\bb n}^{i} \cdot \frac{1}{2} \left[ T^{1/2} \bb\noi^{i} \delta t^{1/2} \right] \bbxi_v^i \dpos } \delta t^{1/2} \\
& = T^{1/2} \sum_i \sum_{uw} \int_V \Avg{ \frac{\delta g}{\delta\bb n}^{i}_u \bb M^{i}_{uw} } \frac{ \Avg{ \bbxi_w^{i} \bbxi_v^i } }{2} \dpos \delta t,
\end{align*}
where we once again used that fields and noises are uncorrelated at same instant $i$.
Using $\Avg{ \bbxi_w^i \bbxi_v^i } = 2 \delta_{wv}$ finally leads us to
\begin{align}
\notag
\Avg{ \int_{t_1}^{t_2} \int_V g(\fld(\pos,t),n(\pos,t)) \bbxi_v(\pos,t) \dpos \D t }
& \approx T^{1/2} \sum_i \sum_{u} \int_V \Avg{ \frac{\delta g}{\delta\bb n}^{i}_u \bb M^{i}_{uv} } \dpos \delta t \\
\label{eq:StratoToRiemann}
& = T^{1/2} \sum_{w} \int_{t_1}^{t_2} \Avg{ \int_V \bb M_{wv} \frac{\delta g}{\delta\bb n}_w \dpos } \D t,
\end{align}
where we returned to the continuum limit in the last line and renamed the dummy variable $u \to w$.
Equation~\eqref{eq:StratoToRiemann}, which translates a stochastic Stratonovich integral into a standard Riemannian integral, is the equivalent of Eq.~(46) in reference~\cite{Cates2022}.

Armed with Eq.~\eqref{eq:StratoToRiemann}, we can now carry out the stochastic integral in Eq.~\eqref{eq:qNoiseMatrixNotation}.
More specifically, setting $g = \sum_u \bb K_u \bb M_{uv}$ and summing Eq.~\eqref{eq:StratoToRiemann} over $v$, we get
\begin{align}
\notag
\Avg{ \int_{t_1}^{t_2} \int_V \bb K^\trp \cdot \bb M \cdot \bbxi \dpos \D t }
& = \sum_{uv} \Avg{ \int_{t_1}^{t_2} \int_V \bb K_u \bb M_{uv} \bbxi_v \dpos \D t } \\
\label{eq:qdotNoisyIntegral1}
& = T^{1/2} \sum_{uvw} \int_{t_1}^{t_2} \Avg{ \int_V \bb M_{wv} \frac{\delta}{\delta\bb n}_{w} \left[ \bb K_u \bb M_{uv} \right] \dpos } \D t.
\end{align}
This result, together with the spurious drift integral in the RHS of Eq.~\eqref{eq:qNoiseMatrixNotation} and spurious drift definition in Eq.~\eqref{eq:SpuriousDrift}, allow us to write
\begin{align}
\notag
\Avg{ \int_{t_1}^{t_2} P_\text{spu}(t) \D t }
& = \frac{T \act}{\mob_\fld} \sum_{uvw} \int_{t_1}^{t_2} \Avg{ \int_V \left\{ \bb M_{wv} \frac{\delta}{\delta\bb n}_{w} \left[ \bb K_u \bb M_{uv} \right] + \bb K_u \bb M_{uv} \frac{\delta}{\delta\bb n}_{w} \bb M_{wv} \right\} \dpos } \D t \\
\notag
& = \frac{T \act}{\mob_\fld} \sum_{uvw} \int_{t_1}^{t_2} \Avg{ \int_V \frac{\delta}{\delta\bb n}_{w} \left[ \bb K_u \bb M_{uv} \bb M_{wv} \right] \dpos } \D t \\
\label{eq:qdotNoisyIntegral2}
& = \frac{T \act}{\mob_\fld} \int_{t_1}^{t_2} \Avg{ \int_V \frac{\delta}{\delta\bb n}^\trp \cdot \left[ \bb L \cdot \bb K \right] \dpos } \D t,
\end{align}
where the last line follows from $\bb L = \bb M \cdot \bb M^\trp$.
Up until here, we have not used the specific form of $\bb K$, and thus the simple form of Eq.~\eqref{eq:qdotNoisyIntegral2} is valid for any column matrix.
However, when we invoke the definitions of $\bb L$ (see Eq.~(6) of the main text) and of $\bb K$ in Eq.~\eqref{eq:MobilityMatrix}, we see that their dot product is very simple, with a single non-zero entry in its last element.
This ultimately takes us to
\begin{equation} \label{eq:qdotNoisyIntegral3}
\Avg{ \int_{t_1}^{t_2} P_\text{spu}(t) \D t }
= \frac{T \act}{\mob_\fld} \int_{t_1}^{t_2} \Avg{ \int_V \frac{\delta}{\delta n} \left[ \mob_n\mob_\fld - \bs\cpl^2 \right] \dpos } \D t = 0.
\end{equation}
The last equality follows from the fact that $\bs\cpl$ does not depend on $n$.

\bibliography{bib_v12}